\documentclass[usenatbib,useAMS,twocolumn]{aastex62}
\usepackage{epsfig,amssymb,amsmath,rotating,graphicx}

\newcommand\chan{{\it Chandra}}
\newcommand\xmm{{\it XMM-Newton}}
\newcommand\rosat{{\it ROSAT}}
\newcommand\asca{{\it ASCA}}

\newcommand\s{{\rm~s}}
\newcommand\ks{{\rm~ks}}

\newcommand\mpc{{\rm~Mpc}}

\newcommand\hz{{\rm~Hz}}

\newcommand\keV{{\rm~keV}}
\newcommand\ev{{\rm~eV}}

\submitjournal{ApJ}
\shorttitle{Rapid X-ray variability in PG~1404$+$226}
\shortauthors{Mallick \& Dewangan}
\begin{document}
\title{Large-amplitude rapid X-ray variability in the narrow-line Seyfert~1 galaxy PG~1404$+$226}
\correspondingauthor{Labani Mallick}
\email{labani@iucaa.in (LM)}

\author{L. Mallick}
\affiliation{Inter-University Center for Astronomy and Astrophysics, Ganeshkhind, Pune 411007, India}

\author{G. C. Dewangan}
\affiliation{Inter-University Center for Astronomy and Astrophysics, Ganeshkhind, Pune 411007, India}


\label{firstpage}

\begin{abstract}
We present the first results from a detailed analysis of a new, long ($\sim100$\ks{}) \xmm{} observation of the narrow-line Seyfert~1 galaxy PG~1404$+$226 which showed a large-amplitude, rapid X-ray variability by a factor of $\sim7$ in $\sim10$\ks{} with an exponential rise and a sharp fall in the count rate. We investigate the origin of the soft X-ray excess emission and rapid X-ray variability in the source through time-resolved spectroscopy and fractional root-mean-squared (rms) spectral modeling. The strong soft X-ray excess below 1\keV{} observed both in the time-averaged and time-resolved spectra is described by the intrinsic disk Comptonization model as well as the relativistic reflection model where the emission is intensive merely in the inner regions ($r_{\rm in}<1.7 r_{\rm g}$) of an ionized accretion disk. We detected no significant UV variability while the soft X-ray excess flux varies together with the primary power-law emission (as $F_{{\rm primary}}\propto F_{{\rm excess}}^{1.54}$), although with a smaller amplitude, as expected in the reflection scenario. The observed X-ray fractional rms spectrum is approximately constant with a drop at $\sim0.6$\keV{} and is described by a non-variable emission line component with the observed energy of $\sim0.6$\keV{} and two variable spectral components: a more variable primary power-law emission and a less variable soft excess emission. Our results suggest the `lamppost geometry' for the primary X-ray emitting hot corona which illuminates the innermost accretion disk due to strong gravity and gives rise to the soft X-ray excess emission.
\end{abstract}

\keywords{accretion, accretion disks --- galaxies: Seyfert --- galaxies: individual: PG~1404$+$226 --- X-rays: galaxies}

\section{Introduction}
The narrow-line Seyfert~1 (NLS1) galaxies, a subclass of active galactic nuclei (AGNs) have been the centre of interest because of their extreme variability in the X-ray band \citep{bo96,le99a,ko00}. The defining properties of this class of AGNs are: Balmer lines with the full width at half-maximum FWHM(H$_\beta$) $<2000$\rm~km~s$^{-1}$ \citep{os85,go89}, strong permitted optical/UV Fe II emission lines \citep{gr99,ve01,bo92} and weaker [OIII] emission $\frac{[OIII]\lambda5007}{H_{\beta}}\leq3$ \citep{go89,os85}. The X-ray spectra of Seyfert galaxies show a power-law like primary continuum which is thought to arise due to thermal Comptonization of the optical/UV seed photons in a corona of hot electrons surrounding the central supermassive black hole (e.g. \citealt{ha91,ha93}). The optical/UV seed photons are thought to arise from an accretion disk \citep{sh73}. However, the interplay between the accretion disk and the hot corona is not well understood. Many type~1 AGNs also show strong `soft X-ray excess' emission over the power-law continuum below $\sim2$\keV{} in their X-ray spectra. The existence of this component ($\sim0.1-2$\keV{}) was discovered around 30~years ago (e.g. \citealt{ar85,si85}), and its origin is still controversial. Initially, it was considered to be the high energy tail of the accretion disk emission (\citealt{ar85,le99b}), but the temperature of the soft X-ray excess is in the range $\sim0.1-0.2$\keV{} which is much higher than the maximum disk temperature expected in AGNs. It was then speculated that the soft X-ray excess could result from the Compton up-scattering of the disk photons in an optically thick, warm plasma (e.g. \citealt{ma98,jan01}). Currently, there are two competing models for the origin of the soft X-ray excess: optically thick, low-temperature Comptonization \citep{ma98,de07,done12} and relativistic reflection from an ionized accretion disk \citep{fa02,cr06,ga14,ma18}. However, these models sometimes give rise to spectral degeneracy because of the presence of multiple spectral components in the energy spectra of NLS1 galaxies \citep{de07,gh16}. One efficient approach to overcome the spectral model degeneracy is to study the root-mean-squared (rms) spectrum which links the energy spectrum with variability and has been successfully applied in a number of AGNs (MCG--6-30-15: \citealt{mi07}, 1H~0707--495: \citealt{fa12}, RX~J1633.3$+$4719: \citealt{ma16}, Ark~120: \citealt{ma17}). Observational evidence for the emission in different bands such as UV, soft and hard X-rays during large variability events may help us to probe the connection between the disk, hot corona and the soft X-ray excess emitting regions. 

In this paper, we investigate the origin of the soft X-ray excess emission, rapid X-ray variability and the disk-corona connection in PG~1404$+$226 with the use of both model dependent and model independent techniques. PG~1404$+$226 is a NLS1 galaxy at a redshift $z=0.098$ with FWHM(H$_\beta$) $\sim800$\rm~km~s$^{-1}$ \citep{wa96}. Previously, the source was observed with \rosat{} (\citealt{um96}), \asca{} (\citealt{le97,va99}), \chan{} (\citealt{da05}) and \xmm{} (\citealt{cr05}). From the \asca{} observation, the $2-10$\keV{} spectrum was found to be quite flat ($\Gamma=1.6\pm0.4$) with flux $F_{\rm 2-10}\sim6.4\times10^{-12}$~erg~cm$^{-2}$~s$^{-1}$ \citep{va99}. The detection of an absorption edge at $\sim1$\keV{} was claimed in previous studies and interpreted as the high-velocity ($0.2-0.3$~$c$) outflow of ionized oxygen \citep{le97}. The source is well-known for its strong soft X-ray excess and large-amplitude X-ray variability on the short timescales (\citealt{um96,da05}). Here we explore the X-ray light curves, time-averaged as well as time-resolved energy spectra, fractional rms variability spectrum and flux--flux plot through a new $\sim100$\ks{} \xmm{} observation of PG~1404$+$226.

We describe the \xmm{} observation and data reduction in Section~\ref{sec:obs}. In Section~\ref{sec:time}, we present the analysis of the X-ray light curves and hardness ratio. In Section~\ref{sec:spec}, we present time-averaged and resolved spectral analyses with the use of both phenomenological and physical models. In Section~\ref{sec:f-f} and \ref{sec:fvar}, we present the flux$-$flux analysis and modeling of the X-ray fractional rms variability spectrum, respectively. Finally, we summarize and discuss our results in Section~~\ref{sec:discussion}. Throughout the paper, the cosmological parameters $H_0=70$\rm ~km~s$^{-1}$~Mpc$^{-1}$, $\Omega_{m}=0.27$, $\Omega_{\Lambda}=0.73$ are adopted. 

\begin{figure}
\centering
\begin{center}
\includegraphics[scale=0.31,angle=-90]{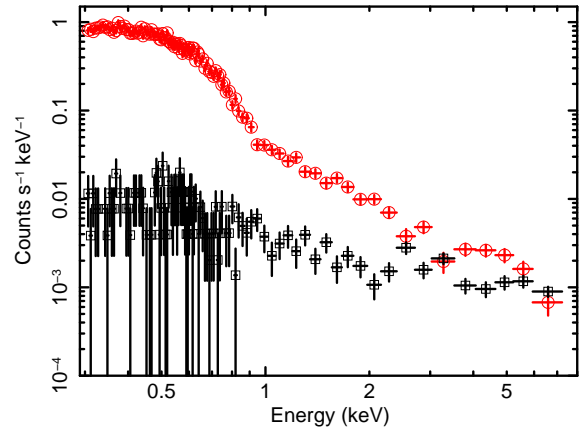}
\caption{The 0.3$-$8\keV{} \xmm{}/EPIC-PN background-subtracted source (in red circle) and background (in black square) spectra of PG~1404$+$226 observed in 2016.}
\label{spec_src_back_gr100}
\end{center}
\end{figure}

\begin{figure*}
\centering
\begin{center}
\includegraphics[scale=0.35,angle=-0]{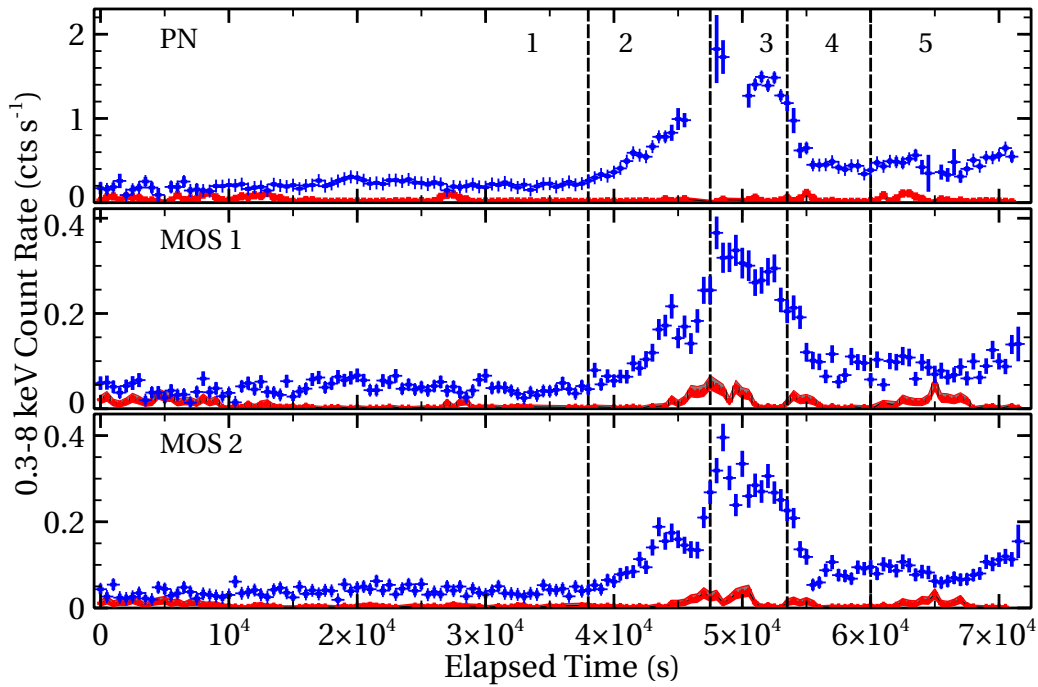}
\caption{The 0.3$-$8\keV{} \xmm{}/EPIC-PN, MOS~1 and MOS~2 background-subtracted, deadtime-corrected light curves of PG~1404$+$226 observed in 2016 with time bins of 500\s{}. The vertical lines indicate the margin between different characteristics of the time series. We show the corresponding background light curves in red.}
\label{lc}
\end{center}
\end{figure*}

\begin{figure*}
\centering
\begin{center}
\includegraphics[scale=0.38,angle=-0]{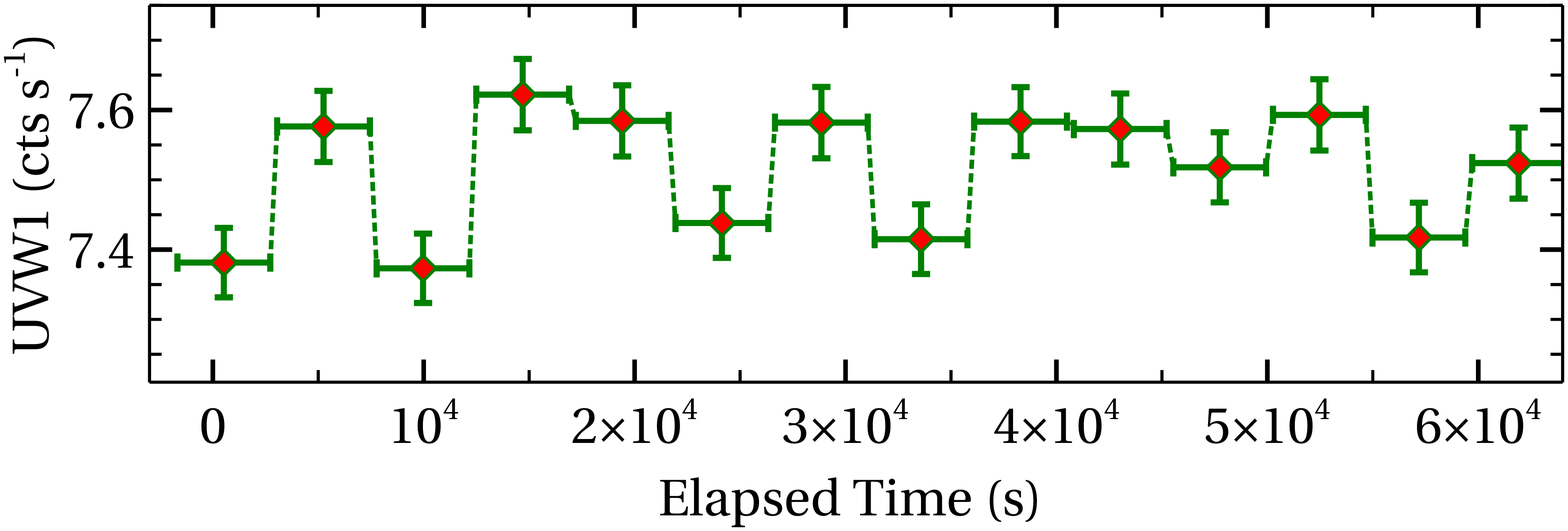}
\caption{The background-subtracted UVW1 count rate of PG~1404$+$226 extracted from the imaging mode OM exposures.}
\label{uv}
\end{center}
\end{figure*}

\begin{figure}
\includegraphics[scale=0.29,angle=-0]{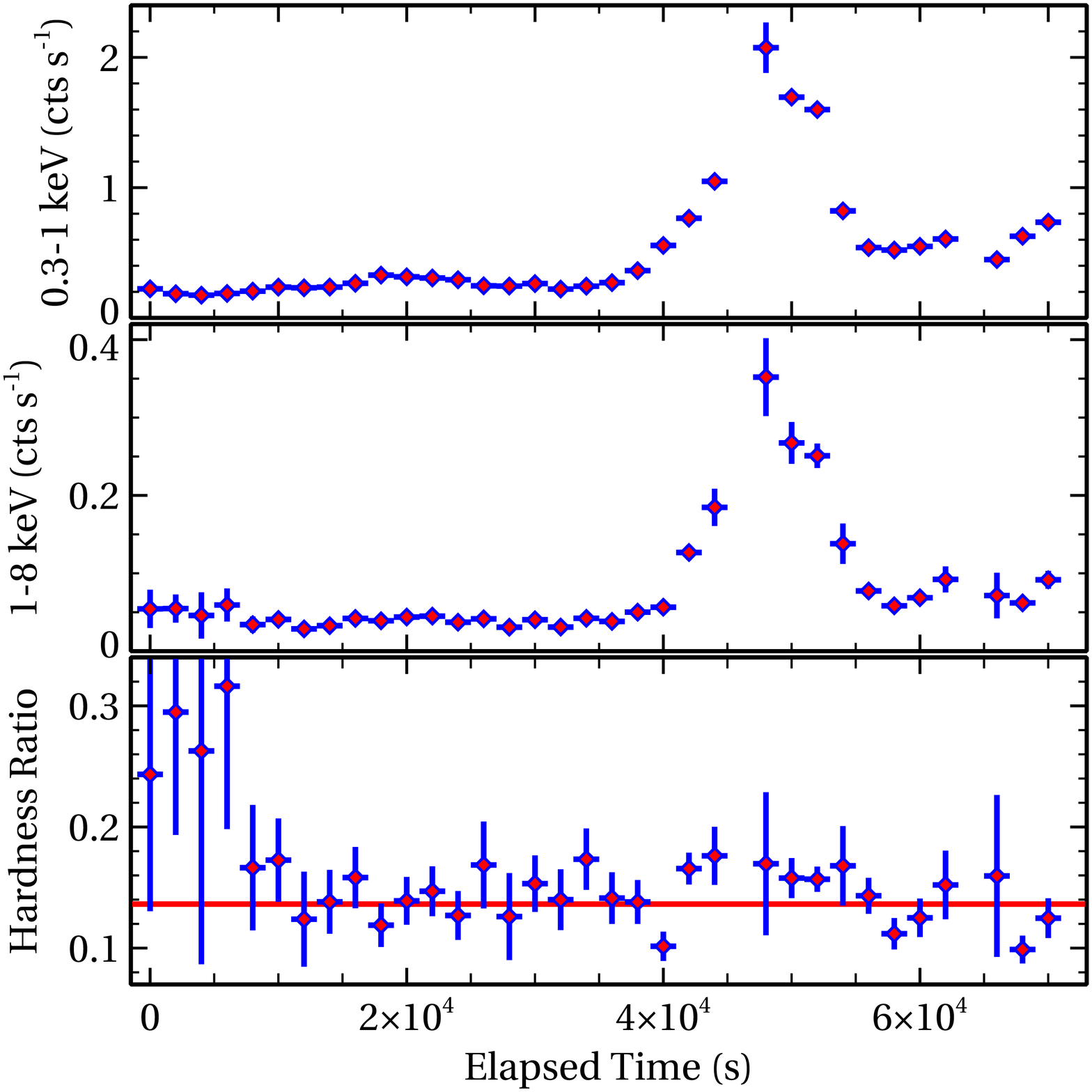}
\caption{The upper and middle panels show the background-subtracted, deadtime-corrected, combined EPIC-PN+MOS light curves in the soft (0.3$-$1\keV{}) and hard (1$-$8\keV{}) bands, respectively. The bottom panel is the corresponding hardness ratio (1$-$8\keV{}/0.3$-$1\keV{}) as a function of the elapsed time, showing spectral variations. The time binning is 2\ks{}.}
\label{hr}
\end{figure}

\section{Observation and Data Reduction}
\label{sec:obs}
We observed PG~1404$+$226 with the \xmm{} telescope \citep{ja01} on 25th January 2016 (Obs.~ID 0763480101) for an exposure time of $\sim100$\ks{}. Here we analyze data from the European Photon Imaging Camera (EPIC-PN; \citealt{st01} and MOS;\citealt{tur01}), Reflection Grating Spectrometer (RGS; \citealt{den01}) and Optical Monitor (OM; \citealt{ma01}) on-board \xmm{}. We processed the raw data with the Scientific Analysis System ({\tt{SAS}}~v.15.0.0) and the most recent (as of 2016 August 2) calibration files. The EPIC-PN and MOS detectors were operated in the large and small window modes, respectively using the thin filter. We processed EPIC-PN and MOS data using {\tt{epproc}} and {\tt{emproc}}, respectively to produce the calibrated photon event files. We checked for the photon pile-up using the task {\tt{epatplot}} and found no pileup in either the PN or MOS data. To filter the processed PN and MOS events, we included unflagged events with ${\tt {pattern}}\le4$ and ${\tt{pattern}}\le12$, respectively. We excluded the proton background flares by generating a {\tt{GTI}} (Good Time Interval) file above 10\keV{} for the full field with {\tt{RATE}}$<3.1$\rm~cts~s$^{-1}$, $1.3$\rm~cts~s$^{-1}$ and $2.1$\rm~cts~s$^{-1}$ for PN, MOS~1 and MOS~2, respectively to obtain the maximum signal-to-noise ratio. It resulted in a filtered duration of $\sim73$\ks{} for both the cleaned EPIC-PN and MOS data. We extracted the PN and MOS source events from a circular region of radii 35~arcsec and 25~arcsec, respectively centered on the source while the background events were extracted from a nearby source-free circular region with a radius of 50~arcsec for both the PN and MOS data. We produced the Redistribution Matrix File ({\tt{rmf}}) and Ancillary Region File ({\tt{arf}}) with the tasks {\tt{rmfgen}} and {\tt{arfgen}}, respectively. We extracted the deadtime-corrected source and background light curves for different energy bands and bin times from the cleaned PN and MOS event files using the task {\tt{epiclccorr}}. We combined the background-subtracted EPIC-PN, MOS~1 and MOS~2 light curves with the {\tt{FTOOLS}} \citep{bl95} task {\tt{lcmath}}. The source count rate was considerably low above 8\keV{}, and therefore we considered only the 0.3$-$8\keV{} band for both the spectral and timing analyses. For spectral analysis, we used only the EPIC-PN data due to their higher signal-to-noise compared to the MOS data. We grouped the average PN spectrum using the {\tt{HEASOFT}}~v.6.19 task {\tt{grppha}} to have a minimum of 50 counts per energy bin. The net count rate estimated for EPIC-PN is ($0.32\pm0.03$)\rm~cts~s$^{-1}$ resulting in a total of $1.65\times10^{4}$ PN counts. Figure~\ref{spec_src_back_gr100} shows the 0.3$-$8\keV{} EPIC-PN background-subtracted source (in red circle) and background (in black square) spectra of PG~1404$+$226.

We processed the RGS data with the {\tt{SAS}} task {\tt{rgsproc}}. The response files were generated using the task {\tt{rgsrmfgen}}. We combined the spectra and response files for two RGS~1+2 using the task {\tt{rgscombine}}. Finally, we grouped the RGS spectral data using the {\tt{grppha}} tool with a minimum of 50 counts per bin. It restricts the applicability of the $\chi^{2}$ statistics.

The Optical Monitor (OM) was operated in the imaging-fast mode using the only UVW1 ($\lambda_{\rm eff}\sim2910~\textrm{\AA}$) filter for a total duration of 94\ks{}. There is a total of 20 UVW1 exposures, and we found that only the last 14 exposures were acquired simultaneously with the filtered EPIC-PN data. We did not use the fast mode OM data due to the presence of a variable background. We processed only the imaging mode OM data with the {\tt{SAS}} task {\tt{omichain}} and obtained the background-subtracted count rate of the source, corrected for coincidence losses.

\section{Timing Analysis: Light curves and Hardness ratio}
\label{sec:time}
We perform the timing analysis of PG~1404$+$226 to investigate the time and energy dependence of variability. Figure~\ref{lc} shows the 0.3$-$8\keV{}, background-subtracted, deadtime-corrected EPIC-PN, MOS~1 and MOS~2 light curves of PG~1404$+$226 with time bins of 500\s{}. The X-ray time series clearly shows a short-term, large-amplitude variability event in which PG~1404$+$226 varied by a factor of $\sim7$ in $\sim10$\ks{} during the 2016 observation. The fractional rms variability amplitude estimated in the 0.3$-$8\keV{} band is $F_{\rm var, X}$=82.5$\pm1.4\%$. The uncertainty on $F_{\rm var}$ was calculated in accordance with \citealt{va03}. Based on the variability pattern, we divided the entire $\sim73$\ks{} light curve into five intervals. Int~1 consists of the first 38\ks{} of the time series and have the lowest flux and moderate fractional rms variability of $F_{\rm var,Int1}$=11.6$\pm2.8\%$. In Int~2, the X-ray flux increases exponentially by a factor of $\sim3$ with fractional rms variability of $F_{\rm var,Int2}$=38.3$\pm2.5\%$. The duration of Int~2 is $\sim10$\ks{}. During Int~3, the source was in the highest flux state with the fractional rms amplitude of $F_{\rm var,Int3}$=9.1$\pm5.1\%$. The source was in the brightest state only for $\sim6$\ks{} and then count rate has started decreasing. In Int~4, the source flux dropped by a factor of $\sim3$ in $\sim6$\ks{} with $F_{\rm var,Int4}$=31.3$\pm3.6\%$. During the end of the observation, the source was moderately variable with $F_{\rm var,Int5}$=8.5$\pm5.9\%$. In Figure~\ref{uv}, we show the UVW1 light curve of PG~1404$+$226 simultaneous with the X-ray light curve. The amplitude of the observed UV variability is only $\sim3\%$ of the mean count rate on timescales of $\sim62$\ks{}. The fractional rms variability amplitude in the UVW1 band is $F_{\rm var,UV}\sim1\%$ which is much less as compared to the X-ray variability. The X-ray and UV variability patterns appear significantly different suggesting lack of any correlation between the X-ray and UV emission at zero time-lag. 

The upper and middle panels of Figure~\ref{hr} show the background-subtracted, deadtime-corrected, combined EPIC-PN+MOS soft (0.3$-$1\keV{}) and hard (1$-$8\keV{}) X-ray light curves, respectively, with time bins of 2\ks{}. The soft band is observed to be brighter than the hard band, however, the variability pattern and amplitude ($F_{\rm var,soft}=88.2\pm1.3\%$ and $F_{\rm var,hard}=88.7\pm5.4\%$) in these two bands are found to be comparable during the observation. The peak-to-trough ratio of the variability amplitude in both the soft and hard bands is of the order of $\sim12$. In the bottom panel of Fig.~\ref{hr}, we have shown the hardness ratio as a function of time. A constant model fitted to the hardness ratio curve provided a statistically poor fit ($\chi^{2}$/d.o.f=46/29), implying the presence of moderate spectral variability and the source became harder at the beginning of the large-amplitude variability. 

\begin{figure}
\includegraphics[scale=0.29,angle=-0]{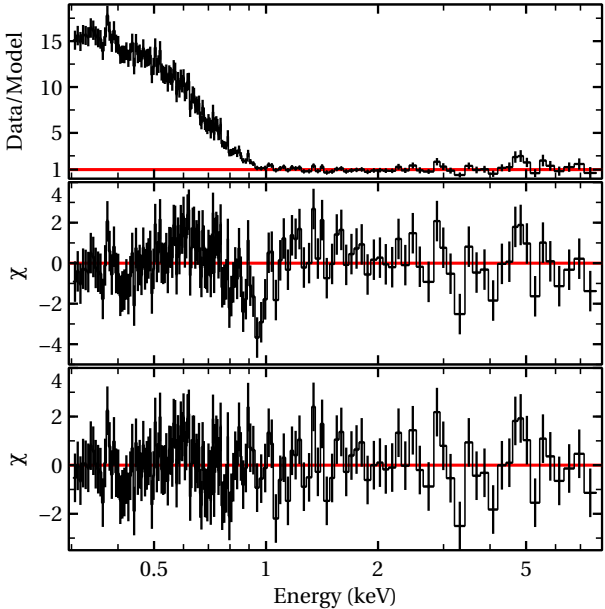}
\caption{Top: The ratio of the EPIC-PN spectral data and absorbed power-law model [{\tt{TBabs$\times$zpowerlw}}] ($\Gamma\sim1.81$) fitted in the 1$-$8\keV{} band and extrapolated to lower energies. Middle: Deviations of the observed PN data from the absorbed blackbody and power-law model [{\tt{TBabs$\times$(zbbody$+$zpowerlw)}}] fitted in the full band (0.3$-$8\keV{}). Bottom: Deviations of the observed PN data from the full band (0.3$-$8\keV{}) model [{\tt{TBabs$\times$WA$\times$(zbbody$+$zpowerlw)}}], showing an excess emission at around 0.6\keV{} in the observer's frame.}
\label{resid}
\end{figure}

\begin{figure*}
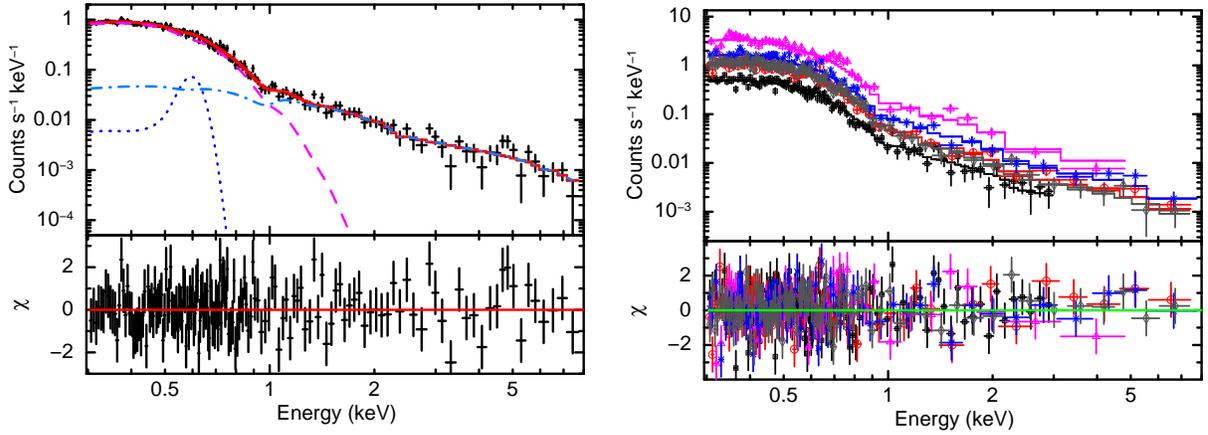

\centering
\begin{center}
\includegraphics[scale=0.31,angle=-90]{fig5a.ps}
\includegraphics[scale=0.31,angle=-90]{fig5b.ps}
\caption{Left: The time-averaged EPIC-PN spectrum, the best-fit model, {\tt{TBabs$\times$WA$\times$(GL$+$zbbody$+$zpowerlw)}} and the deviations of the observed data from the best-fit model (in red). The power-law, blackbody and Gaussian emission components are shown as the dash-dotted, dashed and dotted lines, respectively. Right: The time-resolved EPIC-PN spectra, the best-fit model, {\tt{TBabs$\times$WA$\times$(GL$+$zbbody$+$zpowerlw)}} and the residual spectra. The black squares, red circles, magenta triangles, blue crosses and gray diamonds represent spectral data for Int~1, Int~2, Int~3, Int~4 and Int~5, respectively.}
\end{center}
\label{pheno}
\end{figure*}

\begin{figure*}
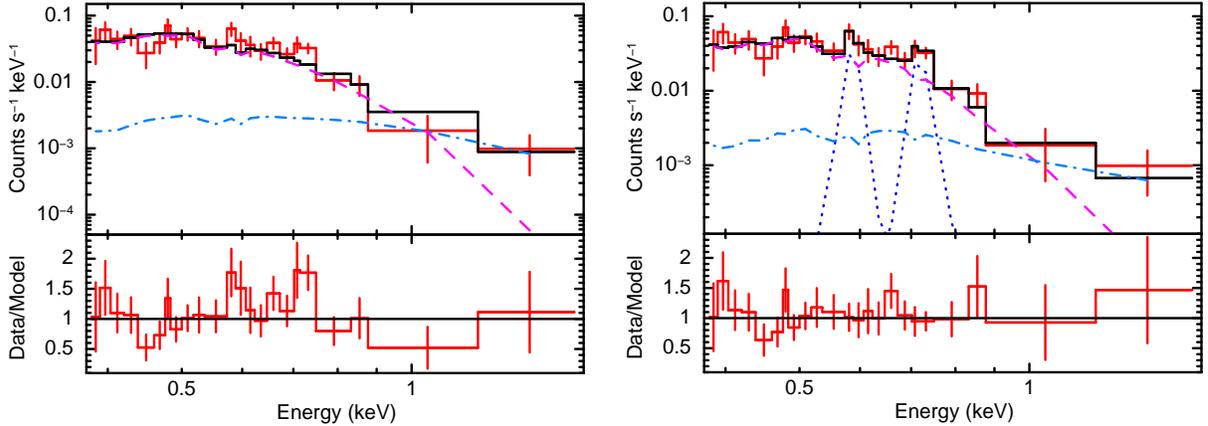

\centering
\begin{center}
\includegraphics[scale=0.31,angle=-90]{fig6a.ps}
\includegraphics[scale=0.31,angle=-90]{fig6b.ps}
\caption{Left: The RGS spectral data, blackbody and power-law models modified by the Galactic absorption and the data-to-model ratio. Right: The RGS spectral data, the best-fit model, {\tt{TBabs$\times$WA$\times$(GL1$+$GL2$+$zbbody$+$zpowerlw)}} and the deviations of the observed data from the best-fit model (in red). The power-law, blackbody and two Gaussian emission components are shown as the dash-dotted, dashed and dotted lines, respectively. The spectra are binned up by a factor of 3 for plotting purposes only.}
\end{center}
\label{rgs}
\end{figure*}

\begin{figure*}
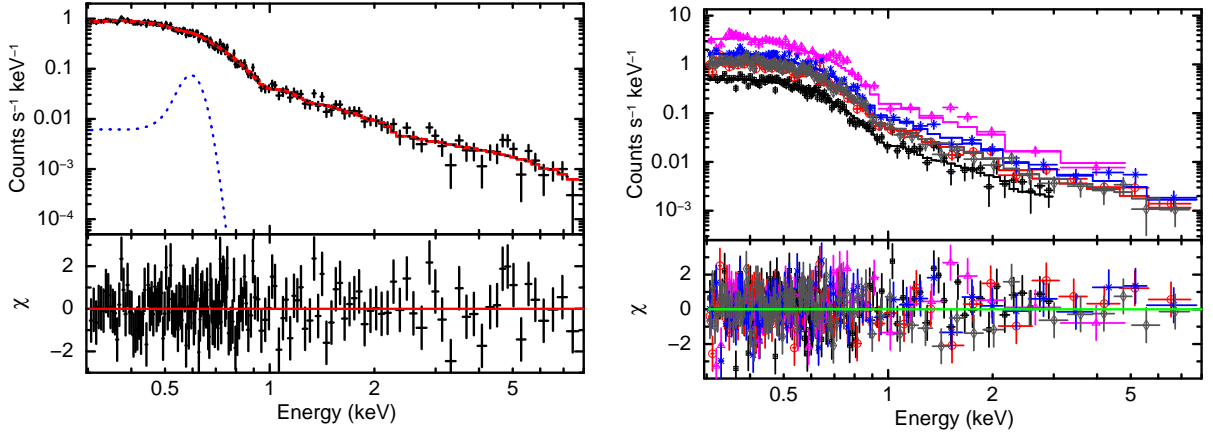

\centering
\begin{center}
\includegraphics[scale=0.31,angle=-90]{fig7a.ps}
\includegraphics[scale=0.31,angle=-90]{fig7b.ps}
\caption{Left: The EPIC-PN mean spectrum, the best-fit absorbed disk Comptonization model, {\tt{TBabs$\times$WA$\times$(GL$+$optxagnf)}} and the deviations of the observed data from the best-fit model. Right: The time-resolved EPIC-PN spectra, the best-fit model, {\tt{TBabs$\times$WA$\times$(GL$+$optxagnf)}} and the residual spectra. The black squares, red circles, magenta triangles, blue crosses and gray diamonds represent spectral data for Int~1, Int~2, Int~3, Int~4 and Int~5, respectively.}
\end{center}
\label{optx}
\end{figure*}

\begin{figure*}
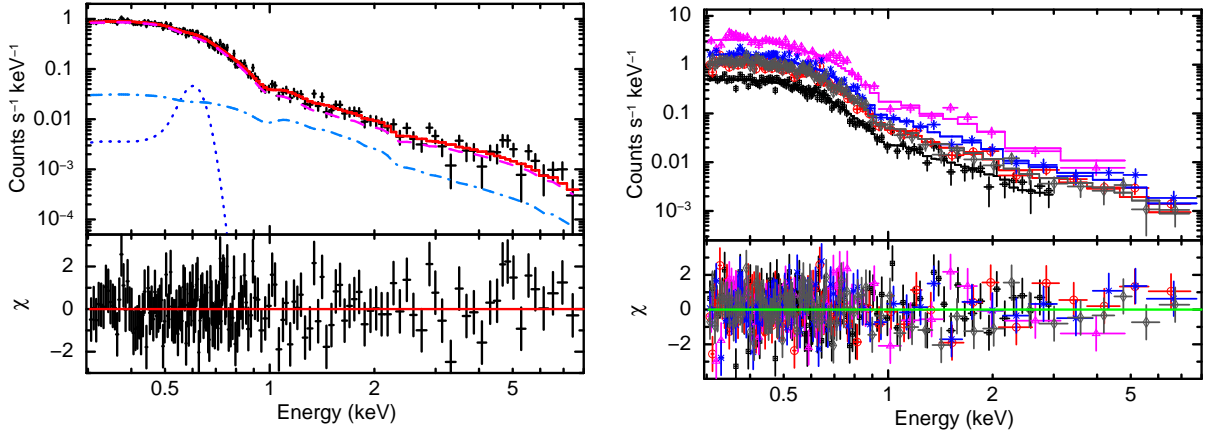

\centering
\begin{center}
\includegraphics[scale=0.31,angle=-90]{fig8a.ps}
\includegraphics[scale=0.31,angle=-90]{fig8b.ps}
\caption{Left: The EPIC-PN mean spectrum, the best-fit absorbed relativistic reflection model (in red) and the deviations of the observed data from the best-fit model, {\tt{TBabs$\times$WA$\times$(GL$+$relconv$\ast$reflionx$+$nthcomp)}}. The primary power-law, relativistic disc reflection and Gaussian emission line components are shown as dash-dotted, dashed and dotted lines, respectively. Right: The 5 time-resolved EPIC-PN spectra, the best-fit model, {\tt{TBabs$\times$WA$\times$(GL$+$relconv$\ast$reflionx$+$nthcomp)}} and the residual spectra. The black squares, red circles, magenta triangles, blue crosses and gray diamonds represent spectral data for Int~1, Int~2, Int~3, Int~4 and Int~5, respectively.}
\end{center}
\label{refl}
\end{figure*}

\section{Spectral Analysis}
\label{sec:spec}
We perform the spectral analysis of PG~1404$+$226 using {\tt{XSPEC}}~v.12.8.2 \citep{ar96}. We employ the $\chi^{2}$ statistics and quote the errors at the $90\%$ confidence limit for a single parameter corresponding to $\Delta\chi^{2}=2.71$ unless otherwise specified. 

\subsection{Phenomenological Model}
\subsubsection{The 0.3$-$8\keV{} EPIC-PN Spectrum}
We begin our spectral analysis by fitting the 1$-$8\keV{} EPIC-PN spectrum using a continuum model ({\tt{zpowerlw}}) multiplied by the Galactic absorption model ({\tt{TBabs}}) using the cross-sections and solar ISM abundances of \citet{wi00}. We fixed the Galactic column density at $N_{\rm H}=2.22\times10^{20}$\rm~cm$^{-2}$ \citep{wil13} after accounting for the effect of molecular hydrogen. This model provided a $\chi^{2}$=70 for 50 degrees of freedom (d.o.f) with $\Gamma\sim1.81$ and can be considered as a good baseline model to describe the hard X-ray emission from the source. Then we extrapolated our 1$-$8~keV{} absorbed power-law model ({\tt{TBabs$\times$zpowerlw}}) down to 0.3\keV{}. This extrapolation reveals the presence of a strong soft X-ray excess emission below 1\keV{} with $\chi^{2}$/d.o.f = 11741/160. We show the ratio of the observed EPIC-PN data and the absorbed power-law model in Figure~\ref{resid}~(top). The fitting of the full band (0.3$-$8\keV{}) data with the absorbed power-law model ({\tt{TBabs$\times$zpowerlw}}) resulted in a poor fit with $\chi^{2}$/d.o.f = 1151.7/158. The residual plot demonstrates a sharp dip in the $0.8-1$\keV{} band and an excess emission below 1\keV{}. Initially, we modeled the soft X-ray excess emission using a simple blackbody model ({\tt{zbbody}}). The addition of the {\tt{zbbody}} model improved the fit statistics to $\chi^{2}$/d.o.f = 230.2/156 ($\Delta\chi^{2}$=$-$921.5 for 2 d.o.f). In {\tt{XSPEC}}, the model reads as {\tt{TBabs$\times$(zbbody$+$zpowerlw)}}. We show the deviations of the observed EPIC-PN data from the absorbed blackbody and power-law model in Fig.~\ref{resid}~(middle). The estimated blackbody temperature $kT_{\rm BB}\sim100$\ev{} is consistent with the temperature of the soft X-ray excess emission observed in Seyfert~1 galaxies and QSOs \citep{cz03,gi04,cr06,pa07}. To model the absorption feature, we have created a warm absorber ({\tt{WA}}) model for PG~1404$+$226 in {\tt{XSTAR}}~v.2.2.1 (last described by \citealt{kb01} and revised in July 2015). The {\tt{XSTAR}} photoionized absorption model has 3 free parameters: column density ($N_{\rm H}$), redshift ($z$) and ionization parameter ($\log\xi$, where $\xi=L/nr^{2}$, $L$ is the source luminosity, $n$ is the hydrogen density and $r$ is the distance between the source and cloud). The inclusion of the warm absorber ({\tt{WA}}) significantly improved the fit statistics from $\chi^{2}$/d.o.f = 230.2/156 to 173.7/154 ($\Delta\chi^{2}$=$-$56.5 for 2 d.o.f). To test the presence of any outflow, we varied the redshift of the absorbing cloud which did not improve the fit statistics. We show the deviations of the observed EPIC-PN data from the model, {\tt{TBabs$\times$WA$\times$(zbbody$+$zpowerlw)}} in Fig.~\ref{resid} (bottom). We notice significant positive residuals at $\sim0.6$\keV{} which may be the signature of an emission feature. To model the emission feature, we added a Gaussian emission line ({\tt{GL}}) which improved the fit statistics to $\chi^{2}$/d.o.f = 154.6/152 ($\Delta\chi^{2}$=$-$19.1 for 2 d.o.f). The centroid energies of the emission line in the observed and rest frames are $\sim0.6$\keV{} and $\sim0.66$\keV{}, respectively. The rest frame $0.66$\keV{} emission feature most likely represents the O~VIII Lyman-$\alpha$ line. The EPIC-PN spectral data, the best-fit model, {\tt{TBabs$\times$WA$\times$(GL$+$zbbody$+$zpowerlw)}} and the deviations of the observed data from the best-fit model are shown in Figure~\ref{pheno}~(left). The best-fit values for the column density, ionization parameter of the warm absorber are $N_{\rm H}=5.2^{+2.9}_{-2.0}\times10^{22}\rm~cm^{-2}$ and $\log(\xi$/erg~cm~s$^{-1})=2.8^{+0.1}_{-0.2}$, respectively.

To search for spectral variability on shorter timescales, we performed time-resolved spectroscopy. First, we generated 5 EPIC-PN spectra from the five intervals defined in Section~\ref{sec:time}. We grouped each spectrum so that we had a minimum of 30 counts per energy bin. The source was hardly detected above 3\keV{} for the lowest flux state corresponding to Int~1. Hence we considered only the $0.3-3$\keV{} energy band for the spectral modelling of Int~1. Then we applied our best-fit mean spectral model to all 5 EPIC-PN spectra. We tied all the parameters except the normalization of the power-law and blackbody components which we set to vary independently. It resulted in a $\chi^{2}$/d.o.f = 464/399, without any strong residuals. If we allow the blackbody temperature and photon index of the power-law to vary, we did not find any significant improvement in the fit with $\chi^{2}$/d.o.f = 446.6/389 ($\Delta\chi^{2}$=$-$17.4 for 10 free parameters). The 5 EPIC-PN spectral data sets, the best-fit model and residuals are shown in Fig.~\ref{pheno}~(right). We list the best-fit spectral model parameters for both the time-averaged and time-resolved spectra in Table~\ref{table1}. 
 
\begin{table*}
 \centering
 \caption{The spectral parameters obtained from the time-averaged and the joint fitting of time-resolved spectra for the best-fit phenomenological model: {\tt{TBabs$\times$WA$\times$(GL$+$zbbody$+$zpowerlw)}}, where {\tt{TBabs}}=Galactic absorption, {\tt{WA}}=warm absorption, {\tt{GL}}=Gaussian emission line, {\tt{zbbody}}=blackbody emission, {\tt{zpowerlw}}=primary power-law emission. Parameters with notations $\ddagger$ and $\ast$ denote fixed and tied values, respectively.}
\begin{center}
\scalebox{0.9}{%
\begin{tabular}{cccccccc} 
\hline
Component & Parameter& Average Spectrum & Int~1 & Int~2 & Int~3 & Int~4 & Int~5 \\   
 & & & $0-38$\ks{} & $38-48$\ks{}&$48-54$\ks{}&$54-60$\ks{}&$60-72$\ks{}  \\  [0.2cm]                               
\hline
{\tt{TBabs}} & $N_{\rm H}$ (10$^{20}$ cm$^{-2}$)$^{a}$ &2.22$^{\ddagger}$&2.22$^{\ast}$ &2.22$^{\ast}$&2.22$^{\ast}$&2.22$^{\ast}$&2.22$^{\ast}$  \\ [0.2cm]

{\tt{WA}} & $N_{\rm H}$ (10$^{22}$cm$^{-2}$)$^{b}$ & 5.2$^{+2.9}_{-2.0}$ & 5.2$^{\ast}$ &5.2$^{\ast}$ &5.2$^{\ast}$ &5.2$^{\ast}$ &5.2$^{\ast}$  \\ [0.2cm]
 & $\log(\xi$/erg cm s$^{-1}$)$^{c}$ & $2.8^{+0.1}_{-0.2}$ & $2.8^{\ast}$ &$2.8^{\ast}$ &$2.8^{\ast}$ &$2.8^{\ast}$ &$2.8^{\ast}$ \\ [0.2cm]
 
 {\tt{GL}} & $E_{\rm rest}$ (keV)$^{d}$ & 0.66$^{+0.02}_{-0.02}$ & 0.66$^{\ast}$ & 0.66$^{\ast}$ & 0.66$^{\ast}$ & 0.66$^{\ast}$ & 0.66$^{\ast}$  \\ [0.2cm] 
 & $E_{\rm obs}$ (keV)$^{e}$ & 0.60$^{+0.02}_{-0.02}$ & 0.6$^{\ast}$ & 0.6$^{\ast}$ & 0.6$^{\ast}$ & 0.6$^{\ast}$ & 0.6$^{\ast}$  \\ [0.2cm] 
 & $\sigma$ (keV)$^{f}$ & 0.01$^{\ddagger}$ & $0.01^{\ast}$ & $0.01^{\ast}$ & $0.01^{\ast}$ & $0.01^{\ast}$ & $0.01^{\ast}$ \\ [0.2cm] 
 
& $A_{\rm GL}$ (10$^{-5}$)$^{g}$ & 1.73$^{+0.66}_{-0.68}$ & $1.73^{\ast}$ & $1.73^{\ast}$ & $1.73^{\ast}$ & $1.73^{\ast}$ & $1.73^{\ast}$ \\ [0.2cm]
      
{\tt{zbbody}} & $kT_{\rm BB}$ (eV)$^{h}$ & 103.1$^{+4.2}_{-2.6}$ & $103.1^{\ast}$ & $103.1^{\ast}$ & $103.1^{\ast}$ & $103.1^{\ast}$ & $103.1^{\ast}$ \\ [0.2cm]
      & $A_{\rm BB}$ (10$^{-5}$)$^{i}$ & 1.51$^{+0.04}_{-0.04}$ & 0.81$^{+0.03}_{-0.02}$ & 1.84$^{+0.07}_{-0.07}$ & 5.59$^{+0.26}_{-0.26}$ &2.76$^{+0.11}_{-0.11}$ & 1.99$^{+0.07}_{-0.07}$ \\ [0.2cm]
      
{\tt{zpowerlw}} & $\Gamma$$^{j}$ & 1.64$^{+0.15}_{-0.14}$ & $1.64^{\ast}$ & $1.64^{\ast}$ & $1.64^{\ast}$ & $1.64^{\ast}$ & $1.64^{\ast}$ \\ [0.2cm]
      & $A_{\rm PL}$ (10$^{-5}$)$^{k}$ & 3.65$^{+0.57}_{-0.51}$ & 2.20$^{+0.27}_{-0.27}$ & 4.38$^{+0.51}_{-0.51}$ & 16.32$^{+2.20}_{-2.20}$ & 7.52$^{+0.95}_{-0.95}$ & 3.75$^{+0.54}_{-0.54}$  \\ [0.2cm]
      
{\tt{FLUX}} & $F_{\rm BB}$(10$^{-13}$)$^{l}$ & 5.9$^{+0.1}_{-0.1}$ & 3.2$^{+0.1}_{-0.1}$ &7.2$^{+0.3}_{-0.3}$ &22.0$^{+1.0}_{-1.0}$ & 10.8$^{+0.4}_{-0.4}$ & 7.8$^{+0.3}_{-0.3}$  \\ [0.2cm]

     & $F_{\rm PL}$(10$^{-13}$)$^{l}$ & 2.1$^{+0.1}_{-0.1}$ & 1.2$^{+0.1}_{-0.1}$ & 2.5$^{+0.3}_{-0.3}$ & 9.2$^{+1.3}_{-1.1}$ & 4.2$^{+0.5}_{-0.5}$ & 2.1$^{+0.3}_{-0.3}$  \\ [0.2cm]

    & $\chi^2$/$\nu$ & 154.6/152 & 464/399 & - & - & - & -   \\
\hline
\end{tabular}}
\end{center}
Notes:~$^a$~Galactic neutral hydrogen column density. $^b$~Column density of the warm absorber (WA). $^c$~Ionization state of the WA, $^d$~Rest frame energy, $^e$~Observed frame energy, $^f$~Emission line width, $^g$~Normalization in units of photons~cm$^{-2}$~s$^{-1}$, $^h$~Blackbody temperature, $^i$~Blackbody normalization, $^j$~Photon index of the primary power-law. $^k$~Power-law normalization in units of photons~cm$^{-2}$~s$^{-1}$~keV$^{-1}$ at 1\keV{}. $^l$~Observed flux in units of erg~cm$^{-2}$~s$^{-1}$.
\label{table1}
\end{table*}

\begin{table*}
 \centering
 \caption{The best-fit spectral model parameters for the absorbed disk Comptonization model: {\tt{TBabs$\times$WA$\times$(GL$+$optxagnf)}}, where {\tt{TBabs}}=Galactic absorption, {\tt{WA}}=warm absorption, {\tt{GL}}=Gaussian emission line, {\tt{optxagnf}}=disk Comptonized continuum. Parameters with notations $\ddagger$ and $\ast$ denote fixed and tied values, respectively. The notation `$p$' in error ranges indicates that the confidence limit did not converge.}
\begin{center}
\scalebox{0.9}{%
\begin{tabular}{cccccccc} 
\hline
Component & Parameter& Average Spectrum & Int~1 & Int~2 & Int~3 & Int~4 & Int~5  \\   
 & & & $0-38$\ks{} & $38-48$\ks{}&$48-54$\ks{}&$54-60$\ks{}&$60-72$\ks{} \\                                 
\hline
{\tt{TBabs}} & $N_{\rm H}$ (10$^{20}$ cm$^{-2}$)$^{a}$ &2.22$^{\ddagger}$&2.22$^{\ast}$ &2.22$^{\ast}$&2.22$^{\ast}$&2.22$^{\ast}$&2.22$^{\ast}$  \\ [0.2cm]

{\tt{WA}} & $N_{\rm H}$ (10$^{22}$cm$^{-2}$)$^{b}$ & 5.5$^{+2.9}_{-2.1}$ & 5.5$^{\ast}$ &5.5$^{\ast}$ &5.5$^{\ast}$ &5.5$^{\ast}$ &5.5$^{\ast}$  \\ [0.2cm]
 & $\log(\xi$/erg cm s$^{-1}$)$^{c}$ & $2.8^{+0.1}_{-0.2}$ & $2.8^{\ast}$ &$2.8^{\ast}$ &$2.8^{\ast}$ &$2.8^{\ast}$ &$2.8^{\ast}$ \\ [0.2cm]
 
{\tt{GL}} & $E_{\rm rest}$ (keV)$^{d}$ & 0.66$^{+0.02}_{-0.02}$ & 0.66$^{\ast}$ & 0.66$^{\ast}$ & 0.66$^{\ast}$ & 0.66$^{\ast}$ & 0.66$^{\ast}$  \\ [0.2cm] 
 & $E_{\rm obs}$ (keV)$^{e}$ & 0.60$^{+0.02}_{-0.02}$ & 0.6$^{\ast}$ & 0.6$^{\ast}$ & 0.6$^{\ast}$ & 0.6$^{\ast}$ & 0.6$^{\ast}$  \\ [0.2cm] 
 & $\sigma$ (keV)$^{f}$ & 0.01$^{\ddagger}$ & $0.01^{\ast}$ & $0.01^{\ast}$ & $0.01^{\ast}$ & $0.01^{\ast}$ & $0.01^{\ast}$ \\ [0.2cm]  
 
& $A_{\rm GL}$ (10$^{-5}$)$^{g}$ & 1.79$^{+0.65}_{-0.69}$ & $1.79^{\ast}$ & $1.79^{\ast}$ & $1.79^{\ast}$ & $1.79^{\ast}$ & $1.79^{\ast}$ \\ [0.2cm]
            
{\tt{optxagnf}} & $M_{\rm BH}$ ($10^{6}M_{\rm \odot}$)$^{h}$ & $4.5$$^{\ddagger}$ & $4.5^{\ast}$& $4.5^{\ast}$& $4.5^{\ast}$& $4.5^{\ast}$&$4.5^{\ast}$ \\ [0.2cm]
                  & $d$ (Mpc)$^{i}$ & 416$^{\ddagger}$ & 416$^{\ast}$& 416$^{\ast}$& 416$^{\ast}$& 416$^{\ast}$& 416$^{\ast}$ \\ [0.2cm]

                  & $\frac{L}{L_{E}}$$^{j}$ & 0.07$^{+0.02}_{-0.01}$ & 0.04$^{+0.001}_{-0.001}$& 0.08$^{+0.002}_{-0.002}$& 0.25$^{+0.01}_{-0.01}$& 0.12$^{+0.004}_{-0.004}$& 0.08$^{+0.002}_{-0.002}$\\ [0.2cm]

                  & $a$$^{k}$ & 0.998$^{\ddagger}$ & 0.998$^{\ast}$& 0.998$^{\ast}$& 0.998$^{\ast}$& 0.998$^{\ast}$& 0.998$^{\ast}$   \\ [0.2cm]
                  & $R_{\rm corona}$ ($R_{\rm g}$)$^{l}$ & 100.0$^{+0p}_{-95.0}$& 100.0$^{\ast}$ & 100.0$^{\ast}$ & 100.0$^{\ast}$ & 100.0$^{\ast}$ & 100.0$^{\ast}$  \\ [0.2cm]

                  & $kT_{\rm SE}$ (eV)$^{m}$ & 104.5$^{+5.4}_{-2.2}$& 104.5$^{\ast}$& 104.5$^{\ast}$& 104.5$^{\ast}$& 104.5$^{\ast}$& 104.5$^{\ast}$  \\ [0.2cm]
                  
                  & $\tau$$^{n}$ &  100.0$^{+0p}_{-47.0}$& 100.0$^{\ast}$ & 100.0$^{\ast}$ & 100.0$^{\ast}$ & 100.0$^{\ast}$ & 100.0$^{\ast}$  \\ [0.2cm]
                  
              & $\Gamma$$^{o}$ & 1.65$^{+0.14}_{-0.14}$ & 1.65$^{\ast}$& 1.65$^{\ast}$ & 1.65$^{\ast}$& 1.65$^{\ast}$& 1.653$^{\ast}$ \\ [0.2cm]
              
            & $f_{\rm PL}$$^{p}$ & 0.46$^{+0.12}_{-0.09}$ & 0.46$^{\ast}$& 0.46$^{\ast}$& 0.46$^{\ast}$ & 0.46$^{\ast}$& 0.46$^{\ast}$ \\ [0.2cm]
            
                  & $\chi^2$/$\nu$ & 154.9/150 & 479.5/404 &- &- &- &-  \\
\hline
\end{tabular}}
\end{center}
Notes:~$^a$~Galactic neutral hydrogen column density. $^b$~Column density of the warm absorber (WA). $^c$~Ionization state of the WA, $^d$~Rest frame energy, $^e$~Observed frame energy, $^f$~Emission line width, $^g$~Normalization in units of photons~cm$^{-2}$~s$^{-1}$, $^h$~SMBH mass, $^i$~Proper distance, $^j$~Eddington ratio, $^k$~SMBH spin, $^l$~Coronal radius, $^m$~Soft excess temperature, $^n$~Optical depth of the warm corona, $^o$~Photon index of the hot coronal emission. $^k$~Fraction of the power below $R_{\rm corona}$ which is emitted in the hard Comptonization component.
\label{table2}
\end{table*}

\begin{table*}
 \centering
\caption{The best-fit spectral model parameters for the absorbed relativistic reflection model: {\tt{TBabs$\times$WA$\times$(GL$+$relconv$\ast$reflionx$+$nthcomp)}}, where {\tt{TBabs}}=Galactic absorption, {\tt{WA}}=warm absorption, {\tt{GL}}=Gaussian emission line, {\tt{relconv$\ast$reflionx}}=relativistic disk reflection, {\tt{nthcomp}}=Illuminating continuum. Parameters with notations $\ddagger$ and $\ast$ denote fixed and tied values, respectively. The notation `$p$' in error ranges indicates that the confidence limit did not converge.}
\begin{center}
\scalebox{0.9}{%
\begin{tabular}{cccccccc} 
\hline
Component & Parameter& Average Spectrum & Int~1 & Int~2 & Int~3 & Int~4 & Int~5  \\   
 & & & $0-38$\ks{} & $38-48$\ks{}&$48-54$\ks{}&$54-60$\ks{}&$60-72$\ks{} \\                                  
\hline
{\tt{TBabs}} & $N_{\rm H}$ (10$^{20}$ cm$^{-2}$)$^{a}$ &2.22$^{\ddagger}$&2.22$^{\ast}$ &2.22$^{\ast}$&2.22$^{\ast}$&2.22$^{\ast}$&2.22$^{\ast}$  \\ [0.2cm]

{\tt{WA}} & $N_{\rm H}$ (10$^{22}$cm$^{-2}$)$^{b}$ & 4.5$^{+3.1}_{-2.0}$ & 4.5$^{\ast}$ &4.5$^{\ast}$ &4.5$^{\ast}$ &4.5$^{\ast}$ &4.5$^{\ast}$  \\ [0.2cm]
 & $\log(\xi_{\rm WA}$/erg cm s$^{-1}$)$^{c}$ & $2.9^{+0.3}_{-0.2}$ & $2.9^{\ast}$ & $2.9^{\ast}$ & $2.9^{\ast}$ & $2.9^{\ast}$ & $2.9^{\ast}$ \\ [0.2cm]

{\tt{GL}} & $E_{\rm rest}$ (keV)$^{d}$ & 0.66$^{+0.02}_{-0.02}$ & 0.66$^{\ast}$ & 0.66$^{\ast}$ & 0.66$^{\ast}$ & 0.66$^{\ast}$ & 0.66$^{\ast}$  \\ [0.2cm] 
 & $E_{\rm obs}$ (keV)$^{e}$ & 0.60$^{+0.02}_{-0.02}$ & 0.6$^{\ast}$ & 0.6$^{\ast}$ & 0.6$^{\ast}$ & 0.6$^{\ast}$ & 0.6$^{\ast}$  \\ [0.2cm] 
 & $\sigma$ (keV)$^{f}$ & 0.01$^{\ddagger}$ & $0.01^{\ast}$ & $0.01^{\ast}$ & $0.01^{\ast}$ & $0.01^{\ast}$ & $0.01^{\ast}$ \\ [0.2cm]  
 
& $A_{\rm GL}$ (10$^{-5}$)$^{g}$ & 1.17$^{+0.75}_{-0.69}$ & $1.17^{\ast}$ & $1.17^{\ast}$ & $1.17^{\ast}$ & $1.17^{\ast}$ & $1.17^{\ast}$ \\ [0.2cm]
            
{\tt{relconv}}    & $q$$^{h}$ & 9.9$^{+0.1p}_{-3.8}$ & 9.9$^{\ast}$& 9.9$^{\ast}$& 9.9$^{\ast}$& 9.9$^{\ast}$& 9.9$^{\ast}$ \\ [0.2cm]

                  & $a$$^{i}$ & 0.998$^{+p}_{-0.006}$ & 0.998$^{\ast}$& 0.998$^{\ast}$& 0.998$^{\ast}$& 0.998$^{\ast}$& 0.998$^{\ast}$ \\ [0.2cm]

      & $R_{\rm in}$($R_{\rm g}$)$^{j}$ & 1.27$^{+0.46}_{-0.03}$ &1.27$^{\ast}$ &1.27$^{\ast}$&1.27$^{\ast}$&1.27$^{\ast}$&1.27$^{\ast}$\\ [0.2cm]
      & $R_{\rm out}$($R_{\rm g}$)$^{k}$ & 1000$^{\ddagger}$ &1000$^{\ast}$&1000$^{\ast}$&1000$^{\ast}$&1000$^{\ast}$&1000$^{\ast}$ \\ [0.2cm]
       & $i$$^{l}$ & 56.8$^{+1.8}_{-12.9}$ & 56.8$^{\ast}$ & 56.8$^{\ast}$& 56.8$^{\ast}$& 56.8$^{\ast}$& 56.8$^{\ast}$  \\ [0.2cm]

{\tt{reflionx}} & $A_{\rm Fe}$$^{m}$ & 3.5$^{+1.2}_{-1.3}$& 3.5$^{\ast}$& 3.5$^{\ast}$& 3.5$^{\ast}$& 3.5$^{\ast}$& 3.5$^{\ast}$  \\ [0.2cm]
                 
               & $\Gamma$$^{n}$ & 2.1$^\ast$& 2.1$^{\ast}$& 2.1$^{\ast}$& 2.1$^{\ast}$& 2.1$^{\ast}$& 2.1$^{\ast}$  \\ [0.2cm]
              
              & $\xi_{\rm disk}$(erg~cm~s$^{-1}$)$^{o}$ & $199^{+29}_{-75}$ & 199$^{\ast}$& 199$^{\ast}$& 199$^{\ast}$& 199$^{\ast}$& 199$^{\ast}$ \\ [0.2cm]
             
              & $A_{\rm REF}$(10$^{-7}$)$^{p}$ & 1.1$^{+1.3}_{-0.3}$ & 0.6$^{+0.02}_{-0.02}$ & 1.4$^{+0.05}_{-0.05}$ & 3.9$^{+0.2}_{-0.2}$ & 2.0$^{+0.1}_{-0.1}$ & 1.5$^{+0.04}_{-0.04}$ \\ [0.2cm] 
                       
{\tt{nthcomp}} & $\Gamma$$^{q}$ & 2.1$^{+0.1}_{-0.1}$& 2.1$^{\ast}$& 2.1$^{\ast}$& 2.1$^{\ast}$& 2.1$^{\ast}$& 2.1$^{\ast}$  \\ [0.2cm]
                  & $A_{\rm NTH}$(10$^{-6}$)$^{r}$ & 12.7$^{+2.9}_{-12.5}$ & 7.9$^{+3.7}_{-3.7}$ & 9.2$^{+7.8}_{-7.8}$ & 85.7$^{+33.7}_{-33.7}$ & 29.1$^{+14.5}_{-14.6}$ & $<6.2$ \\ [0.2cm]

                  & $\chi^2$/$\nu$ & 157.1/147 & 465.2/399 &-&-&-&- \\
\hline
\end{tabular}}
\end{center}
Notes:~$^a$~Galactic neutral hydrogen column density. $^b$~Column density of the warm absorber (WA). $^c$~Ionization state of the WA, $^d$~Rest frame energy, $^e$~Observed frame energy, $^f$~Emission line width, $^g$~Normalization in units of photons~cm$^{-2}$~s$^{-1}$, $^h$~Emissivity index, $^i$~SMBH spin, $^j$~Inner disk radius, $^k$~Outer disk radius, $^l$~Disk inclination angle in degree, $^m$~Iron abundance (solar), $^n$~Photon index of the relativistic reflection component, $^o$~Disk ionization parameter, $^p$~Normalization of the relativistic reflection component, $^q$~Photon index of the illuminating continuum. $^r$~Normalization of the illuminating continuum.
\label{table3}
\end{table*}
 
\subsubsection{The 0.38$-$1.8\keV{} RGS Spectrum}
To confirm the presence of the warm absorption or emission features, we performed a detailed spectral analysis of the high-resolution RGS data. Initially, we used a continuum model similar to that obtained from the EPIC-PN data, i.e. the sum of a power-law and a blackbody. To account for the cross-calibration uncertainties, we multiplied a constant component. All the parameter values are fixed to the best-fit EPIC-PN value since the RGS data ($0.38-1.8$\keV{}) alone cannot constrain them. In {\tt{XSPEC}}, the model reads as {\tt{constant$\times$TBabs$\times$(zbbody$+$zpowerlw)}}. This model provided a poor fit with $\Delta\chi^{2}$=64 for 46 d.o.f. The RGS spectral data, the fitted continuum model {\tt{constant$\times$TBabs$\times$(zbbody$+$zpowerlw)}} and the deviations of the observed data from the model are shown in Figure~\ref{rgs}~(left). The residual plot shows an absorption feature at $\sim0.9-1.1$\keV{} and two emission features at $\sim0.6$\keV{} and $\sim0.7$\keV{} in the observer's frame. We added two narrow Gaussian emission lines to model these two emission features and a warm absorber ({\tt{WA}}) model to fit the absorption feature, which improved the fit statistics by $\Delta\chi^{2}=-30$ for 6 d.o.f with $\chi^{2}$/d.o.f = 34/40. If we allow the redshift of the {\tt{WA}} model to vary, we did not find any significant improvement in the fit statistics. The rest-frame energies of the emission lines are $0.65^{+0.01}_{-0.01}$\keV{} and $0.78^{+0.01}_{-0.01}$\keV{}, which can be attributed to the O~VIII Lyman-$\alpha$ and Lyman-$\beta$, respectively. The best-fit values for the derived WA parameters are $N_{\rm H}=1.6^{+2.1}_{-1.1}\times10^{23}\rm~cm^{-2}$ and $\log(\xi$/erg~cm~s$^{-1})=2.4^{+1.2}_{-0.3}$. The RGS spectum, the best-fit model, {\tt{constant$\times$TBabs$\times$WA$\times$(GL1$+$GL2$+$zbbody$+$zpowerlw)}} and the deviations of the observed data from the best-fit model are shown in Fig.~\ref{rgs}~(right).

\subsection{Physical Model}
To examine the origin of the soft X-ray excess emission, we have tested two different physical models$-$ thermal Comptonization in an optically thick, warm medium and relativistic reflection from an ionized accretion disk.
First, we have used the intrinsic disk Comptonization model ({\tt{optxagnf}}; \citealt{done12}) which assumes that the gravitational energy released in the disk is radiated as a blackbody emission down to the coronal radius, $R_{\rm corona}$. Inside the coronal radius, the gravitational energy is dissipated to produce the soft X-ray excess component in an optically thick, warm ($kT_{\rm SE}\sim0.2$\keV{}) corona and the hard X-ray power-law tail in an optically thin, hot ($kT_{\rm e}\sim100$\keV{}) corona above the disk. Thus, this model represents an energetically self-consistent model. The four parameters which determine the normalization of the model are the following: black hole mass ($M_{\rm BH}$), dimensionless spin parameter ($a$), Eddington ratio ($\frac{L}{L_{E}}$) and proper distance ($d$). We fitted the 0.3$-$8\keV{} EPIC-PN time-averaged spectrum with the {\tt{optxagnf}} model modified by the Galactic absorption ({\tt{TBabs}}). We fixed the black hole mass, outer disk radius and proper distance at $4.5\times10^6 M_{\rm \odot}$ \citep{ka00,wa01}, 1000$R_{\rm g}$ and 416\mpc{}, respectively. We assumed a maximally rotating black hole as concluded by \citet{cr05} and fixed the spin parameter at $a=0.998$. This model resulted in a statistically unacceptable fit with $\chi^{2}$/d.o.f = 234.9/154, a sharp dip at $\sim0.9$\keV{} and an emission feature at $\sim0.6$\keV{} in the residual spectrum. As before, we used the warm absorber ({\tt{WA}}) model which significantly improved the fit statistics to $\chi^{2}$/d.o.f = 175/152 ($\Delta\chi^{2}$=$-$59.9 for 2 d.o.f). The addition of the Gaussian emission line ({\tt{GL}}) provided a statistically acceptable fit with $\chi^{2}$/d.o.f = 154.9/150 ($\Delta\chi^{2}$=$-$20.1 for 2 d.o.f). The EPIC-PN mean spectrum, the best-fit absorbed disk Comptonization model, {\tt{TBabs$\times$WA$\times$(GL$+$optxagnf)}} and the residuals are shown in Figure~\ref{optx}~(left). The best-fit values for the Eddington rate, coronal radius, electron temperature, optical depth and spectral index are $\frac{L}{L_{E}}=0.07^{+0.02}_{-0.01}$, $R_{\rm corona}=100.0^{+0p}_{-95.0}R_{\rm g}$, $kT_{\rm SE}= 104.5^{+5.4}_{-2.2}$\ev{}, $\tau=100.0^{+0p}_{-47.0}$ and $\Gamma= 1.65^{+0.14}_{-0.14}$, respectively. Then we jointly fitted the five time-resolved spectral data sets with the absorbed disk Comptonization model and kept all the parameters tied to their mean spectral best-fit values except the Eddington ratio. It provided a $\chi^{2}$/d.o.f = 479.5/404, and we did not notice any strong feature in the residual spectra. The 5 EPIC-PN spectral data sets, the best-fit disk Comptonization model and residuals are shown in Fig.~\ref{optx}~(right). The best-fit spectral model parameters for both the time-averaged and time-resolved spectra are listed in Table~\ref{table2}.

The soft X-ray excess emission may also arise due to the relativistic reflection from an ionized accretion disk \citep{fa02,cr06,ga14}. Hence we modeled the soft X-ray excess using the reflection model ({\tt{reflionx}}; \citealt{ro05}) convolved with the {\tt{relconv}} model \citep{ga14} which blurs the spectrum due to general relativistic effects close to the SMBH. We fitted the 0.3$-$8\keV{} EPIC-PN mean spectrum with the thermally Comptonized primary continuum ({\tt{nthcomp}}; \citealt{zd96}) and relativistic reflection model ({\tt{relconv$\ast$reflionx}}) after correcting for the Galactic absorption ({\tt{TBabs}}). The electron temperature of the hot plasma and the disk blackbody seed photon temperature in the {\tt{nthcomp}} model were fixed at 100\keV{} and 50\ev{}, respectively. The parameters of the {\tt{reflionx}} model are iron abundance ($A_{\rm Fe}$), ionization parameter ($\xi_{\rm disk}=4\pi F/n$, $F$ is the total illuminating flux, $n$ is hydrogen density), normalization ($A_{\rm REF}$) of the reflected spectrum and photon index ($\Gamma$) of the incident power-law. The convolution model {\tt{relconv}} has five free parameters: emissivity index ($q$, where emissivity of the reflected emission is defined by $\epsilon\propto R^{-q}$), inner disk radius ($R_{\rm in}$), outer disk radius ($R_{\rm out}$), black hole spin ($a$) and disk inclination angle ($i^{\circ}$). We fixed the outer disk radius at $R_{\rm out}=1000r_{\rm g}$. In {\tt{XSPEC}}, the 0.3$-$8\keV{} model reads as {\tt{TBabs$\times$(relconv$\ast$reflionx$+$nthcomp)}} which provided a reasonably good fit with $\chi^{2}$/d.o.f = 180.8/151. However, the residual spectrum shows an absorption dip at $\sim0.9$\keV{} and an excess emission at $\sim0.6$\keV{}. As before, we fitted the absorption dip with the ionized absorption ({\tt{WA}}). The multiplication of the warm absorber model improved the fit statistics to $\chi^{2}$/d.o.f = 165.1/149 ($\Delta\chi^{2}$=$-$15.7 for 2 d.o.f). To model the emission feature at $\sim0.6$\keV{}, we added a Gaussian emission line ({\tt{GL}}), which provided an improvement in the fit statistics with $\chi^{2}$/d.o.f = 157.1/147 ($\Delta\chi^{2}$=$-$8 for 2 d.o.f). The EPIC-PN mean spectrum, the best-fit absorbed relativistic reflection model, {\tt{TBabs$\times$WA$\times$(GL$+$relconv$\ast$reflionx$+$nthcomp)}} and the residuals are shown in Figure~\ref{refl} (left). The best-fit values for the emissivity index, inner disk radius, disk ionization parameter, black hole spin, disk inclination angle and spectral index of the incident continuum are $q=9.9^{+0.1p}_{-3.8}$, $R_{\rm in}=1.27^{+0.46}_{-0.03}R_{\rm g}$, $\xi= 199^{+29}_{-75}$~erg~cm~s$^{-1}$, $a=0.998^{+p}_{-0.006}$, $i^{\circ}=56.8^{+1.8}_{-12.9}$ and $\Gamma= 2.1^{+0.1}_{-0.1}$, respectively. We also fitted the five time-resolved spectra jointly with the absorbed relativistic reflection model and tied every parameter to its mean spectral best-fit value except the normalization ($A_{\rm REF}$) of the reflection component. This provided an unacceptable fit with $\chi^{2}$/d.o.f = 488.4/404. We then set the normalization ($A_{\rm NTH}$) of the illuminating continuum to vary between the five spectra and obtain a noticeable improvement in the fitting with $\chi^{2}$/d.o.f = 465.2/399 ($\Delta\chi^{2}$=$-$23.2 for 5 d.o.f). If we leave the spectral index ($\Gamma$) of the incident continuum to vary, we did not get any significant improvement in the fitting. We summarize the best-fit spectral model parameters for both time-averaged and time-resolved spectra in Table~\ref{table3}. The EPIC-PN spectral data sets, the best-fit absorbed relativistic reflection model and residuals are shown in Fig.~\ref{refl}~(right).

\begin{figure*}
\centering
\begin{center}
\includegraphics[width=0.3\textwidth,angle=-0]{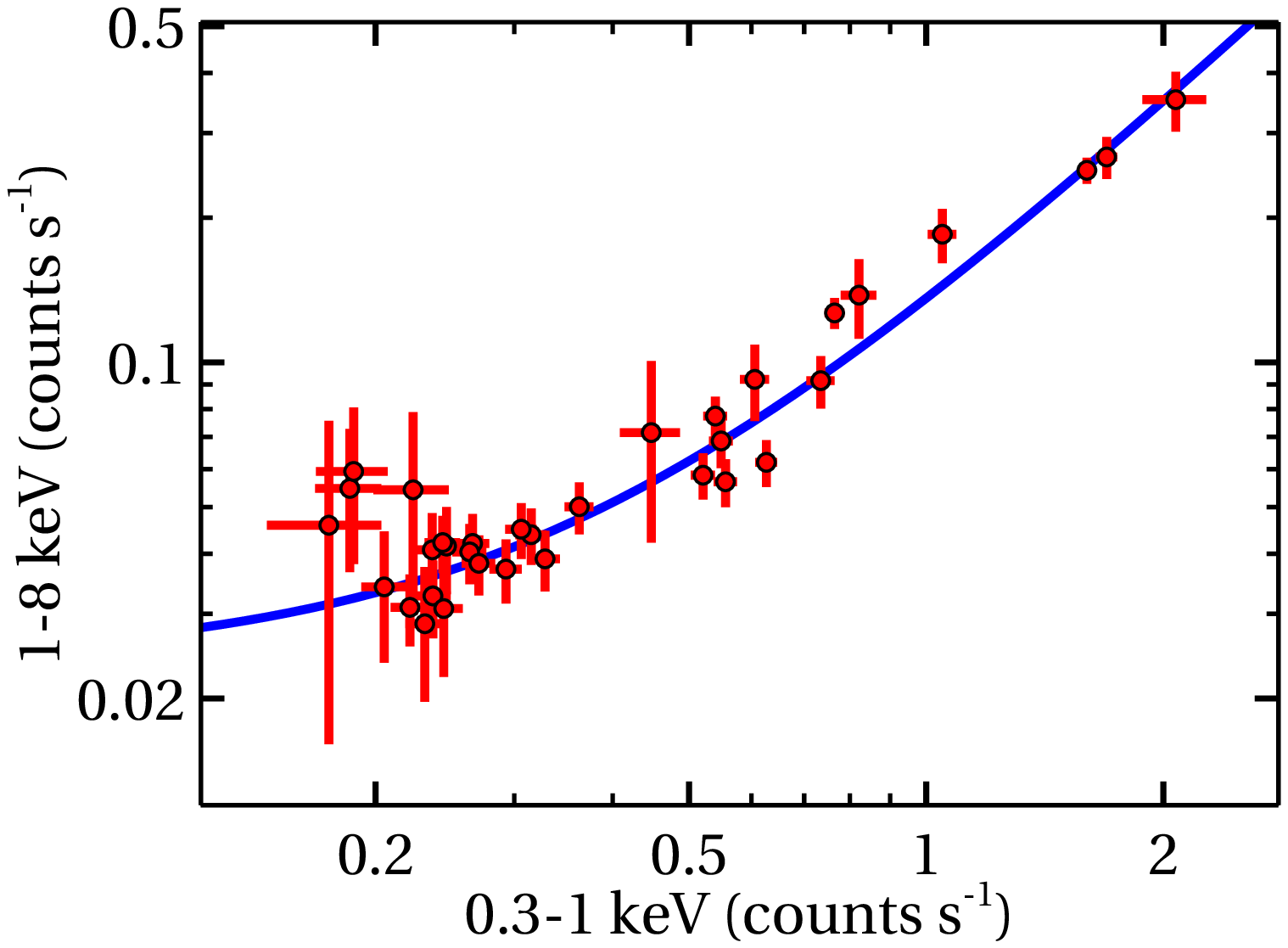}
\includegraphics[width=0.3\textwidth,angle=-0]{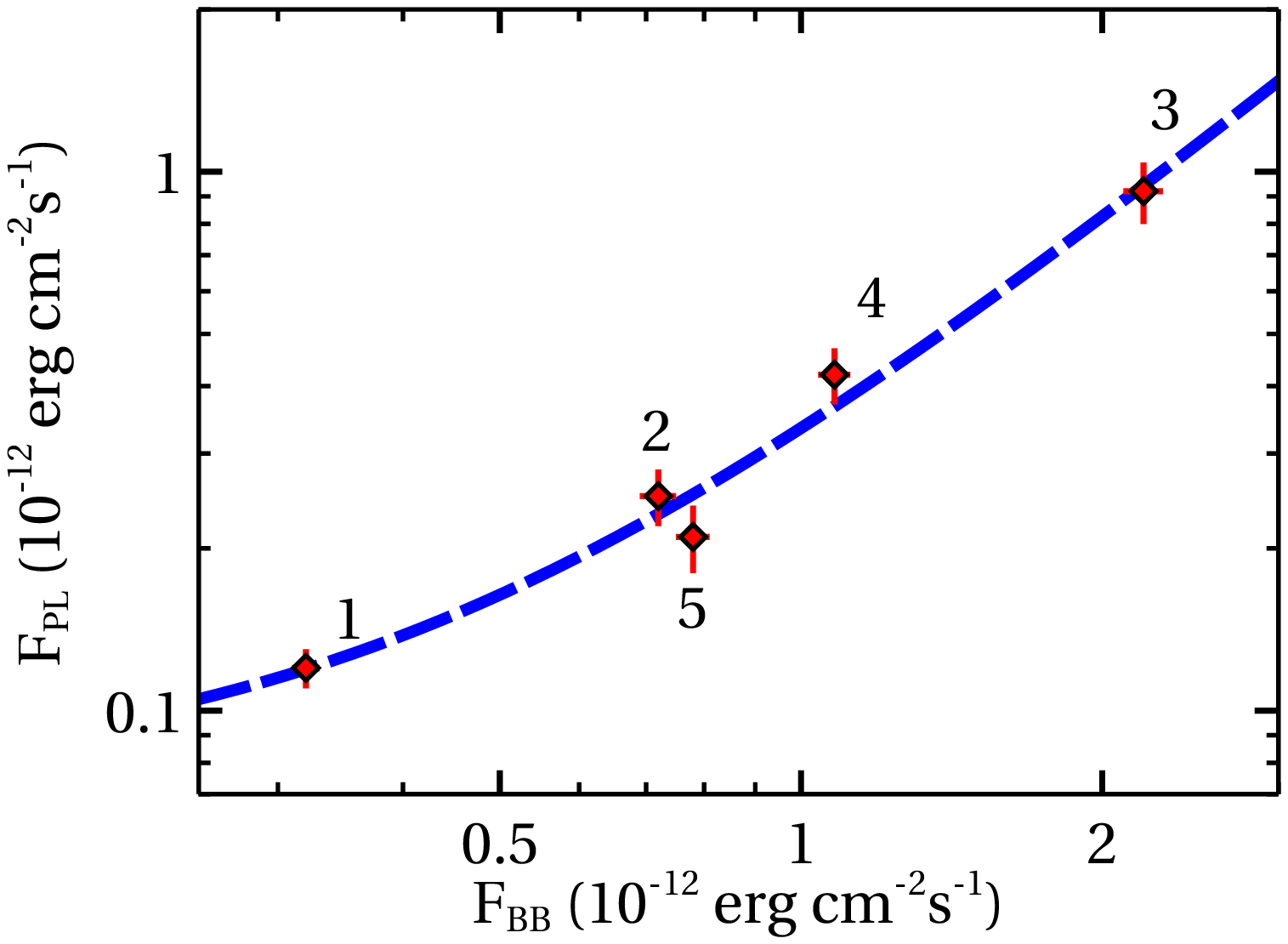}
\includegraphics[width=0.3\textwidth,angle=-0]{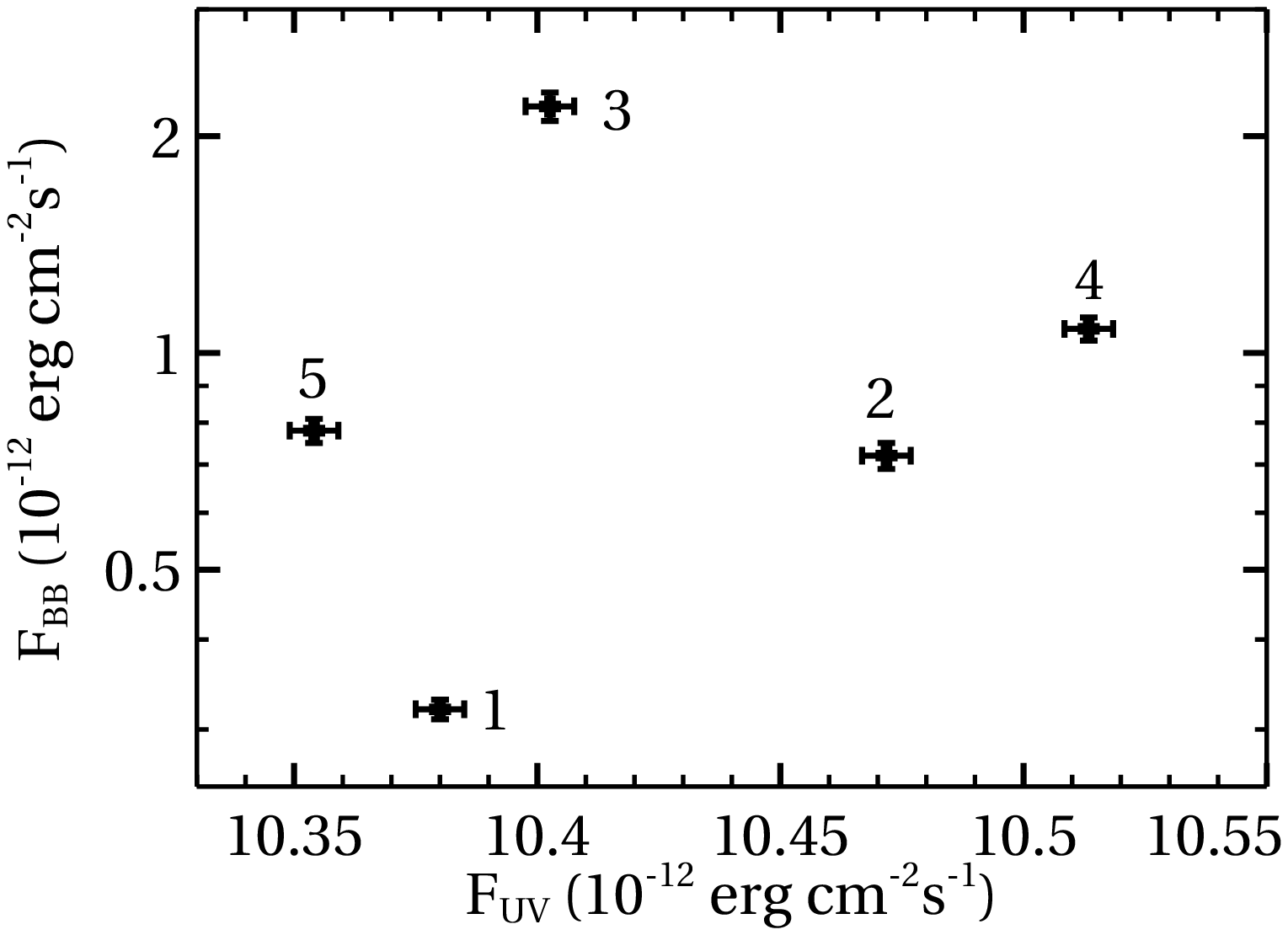}
\caption{Left: The 1$-$8\keV{} count rate is plotted as a function of the 0.3$-$1\keV{} count rate with time bin size of 2\ks{}. The solid line represents the best-fit power-law plus constant (PLC) model. Middle: The unabsorbed primary power-law flux vs blackbody flux in the full band (0.3$-$8\keV{}) obtained from 5 different intervals (marked as 1,2,3,4 and 5). The dashed line shows the best-fit PLC model fitted to the data. Right: The soft X-ray excess flux as a function of the UVW1 flux, implying lack of correlation between the UV and soft X-ray bands.}
\label{fig6}
\end{center}
\end{figure*}
 
\begin{figure*}
\begin{center}
\includegraphics[scale=0.5,angle=-0]{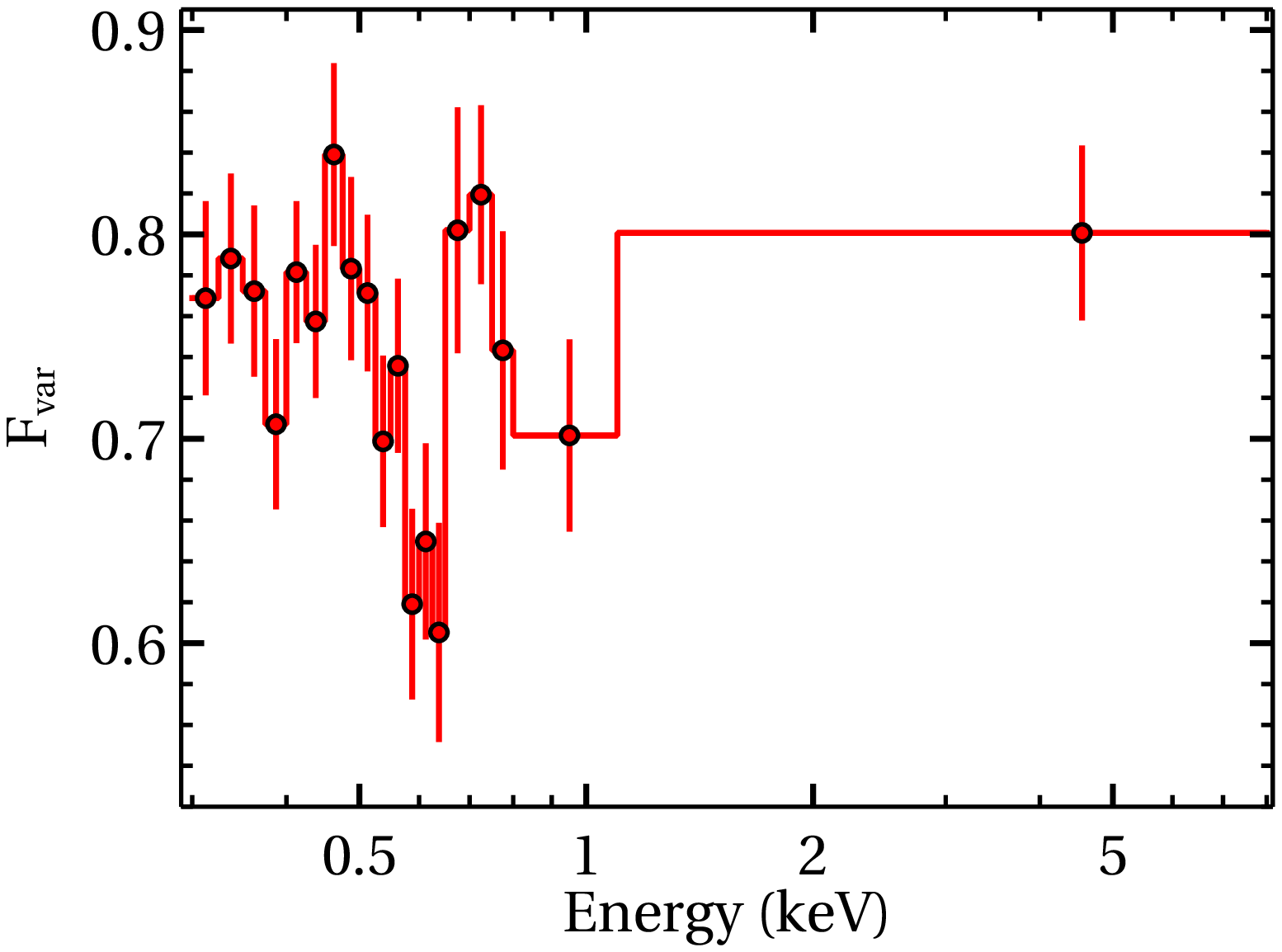}
\includegraphics[scale=0.5,angle=-0]{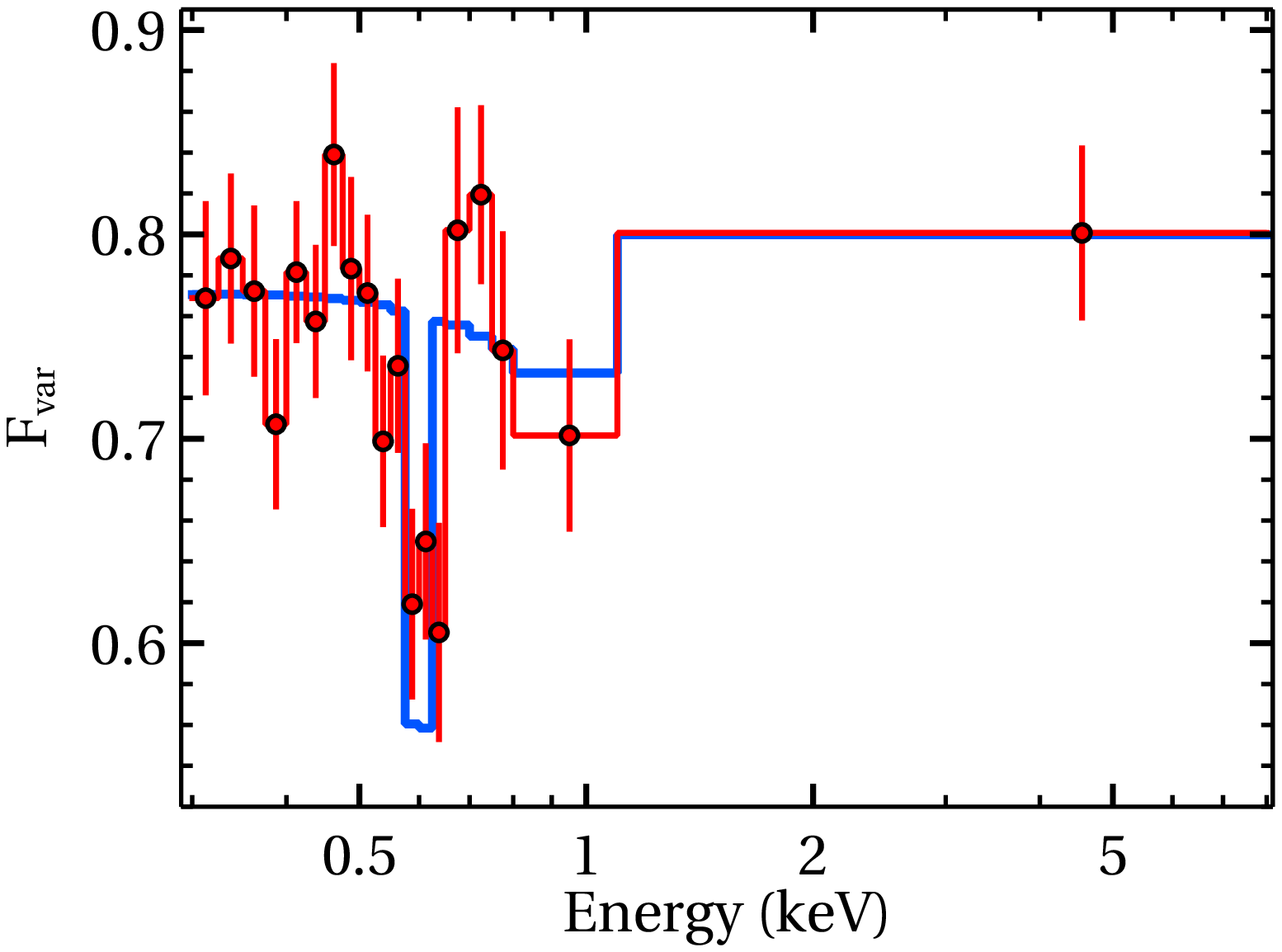}
\caption{Left: The combined EPIC-PN$+$MOS, 0.3$-$8\keV{} fractional rms spectrum of PG~1404$+$226. Right: The solid blue line represents the best-fit `two-component phenomenological' model in which the 0.6\keV{} emission line ({\tt{GL}}) is constant, and both the soft X-ray excess ({\tt{zbbody}}) and primary power-law emission ({\tt{zpowerlw}}) are variable in normalization and positively correlated with each other.}
\label{fig7a}
\end{center}
\end{figure*}

\section{Flux$-$Flux Analysis}
\label{sec:f-f}
We perform the flux$-$flux analysis which is a model-independent approach to distinguish between the main components responsible for the observed spectral variability and was pioneered by \citet{ch01} and \citet{ta03}. Based on our X-ray spectral modeling, we identified the 0.3$-$1\keV{} and 1$-$8\keV{} energy bands as representatives of the soft X-ray excess and primary power-law emission, respectively. Then, we constructed the 0.3$-$1 vs 1$-$8\keV{} flux$-$flux plot which is shown in Figure~\ref{fig6}~(Left). The mean count rate in the soft and hard bands are $0.52\pm0.04$\rm~counts~s$^{-1}$ and $0.08\pm0.02$\rm~counts~s$^{-1}$, respectively. We begin our analysis by fitting the flux$-$flux plot with a linear relation of the form, $y=ax+b$, where $y$ and $x$ represent the 1$-$8\keV{} and 0.3$-$1\keV{} band count rates, respectively. The straight line model provided a statistically unacceptable fit with $\chi^{2}$/d.o.f = 59/32 and implied that the immanent relationship between the soft X-ray excess and primary power-law emission is not linear. Therefore, we fit the flux$-$flux plot with a power-law plus constant (PLC) model of the form, $y=\alpha x^{\beta}+c$ (where $y\equiv1-8\keV{}$ and $x\equiv0.3-1\keV{}$ count rates) following the approach of \citet{ka15}. The PLC model improved the fit statistics to $\chi^{2}$/d.o.f = 37.4/31 and explained the flux$-$flux plot quite well. We show the best-fit PLC model as the solid line in Fig.~\ref{fig6}~(Left). The best-fit power-law normalization, slope and constant values are $\alpha=0.11^{+0.01}_{-0.01}$\rm~counts~s$^{-1}$, $\beta=1.54^{+0.2}_{-0.2}$ and $c=0.02^{+0.006}_{-0.007}$\rm~counts~s$^{-1}$, respectively. The PLC best-fit slope is greater than unity, which indicates the presence of intrinsic variability in the source. The detection of the positive `$c$'-value in the flux$-$flux plot implies that there exists a distinct spectral component which is less variable as compared to the primary X-ray continuum and contributes $\sim25\%$ of the 1$-$8\keV{} count rate at the mean flux level over the observed $\sim20$~hr timescales. To investigate this issue further, we computed the unabsorbed (without the Galactic and intrinsic absorption) primary continuum and soft X-ray excess flux in the full band (0.3$-$8\keV{}) for all five intervals using {\tt{XSPEC}} convolution model {\tt{cflux}} and plotted the intrinsic primary power-law flux as a function of the soft X-ray excess flux (the middle panel of Fig.~\ref{fig6}). The best-fit normalization, slope and constant  parameters, obtained by fitting the $F_{\rm PL}$ vs $F_{\rm BB}$ plot with a PLC model, are $\alpha_{\rm mod}=0.26^{+0.08}_{-0.07}(\times10^{-12})$\rm~erg~cm$^{-2}$s$^{-1}$, $\beta_{\rm mod}=1.53^{+0.43}_{-0.45}$ and $c_{\rm mod}=0.07^{+0.03}_{-0.06}(\times10^{-12})$\rm~erg~cm$^{-2}$s$^{-1}$, respectively. Interestingly, we found steeping in the $F_{\rm PL}$ vs $F_{\rm BB}$ plot with an apparent positive constant which is in agreement with the 0.3$-$1 vs 1$-$8\keV{} flux$-$flux plot. Our flux$-$flux analysis suggests that the primary power-law and soft X-ray excess emission are well correlated with each other, although they vary in a non-linear fashion on the observed timescale. We also investigated the variability relation between the UV and soft X-ray excess emission in PG~1404$+$226. Fig.~\ref{fig6}~(Right) shows the variation of the soft X-ray excess flux as a function of the UVW1 flux, which indicates no significant correlation between the UV and soft X-ray excess emission from PG~1404$+$226.

\begin{figure*}
\centering
\begin{center}
\includegraphics[scale=0.5,angle=-0]{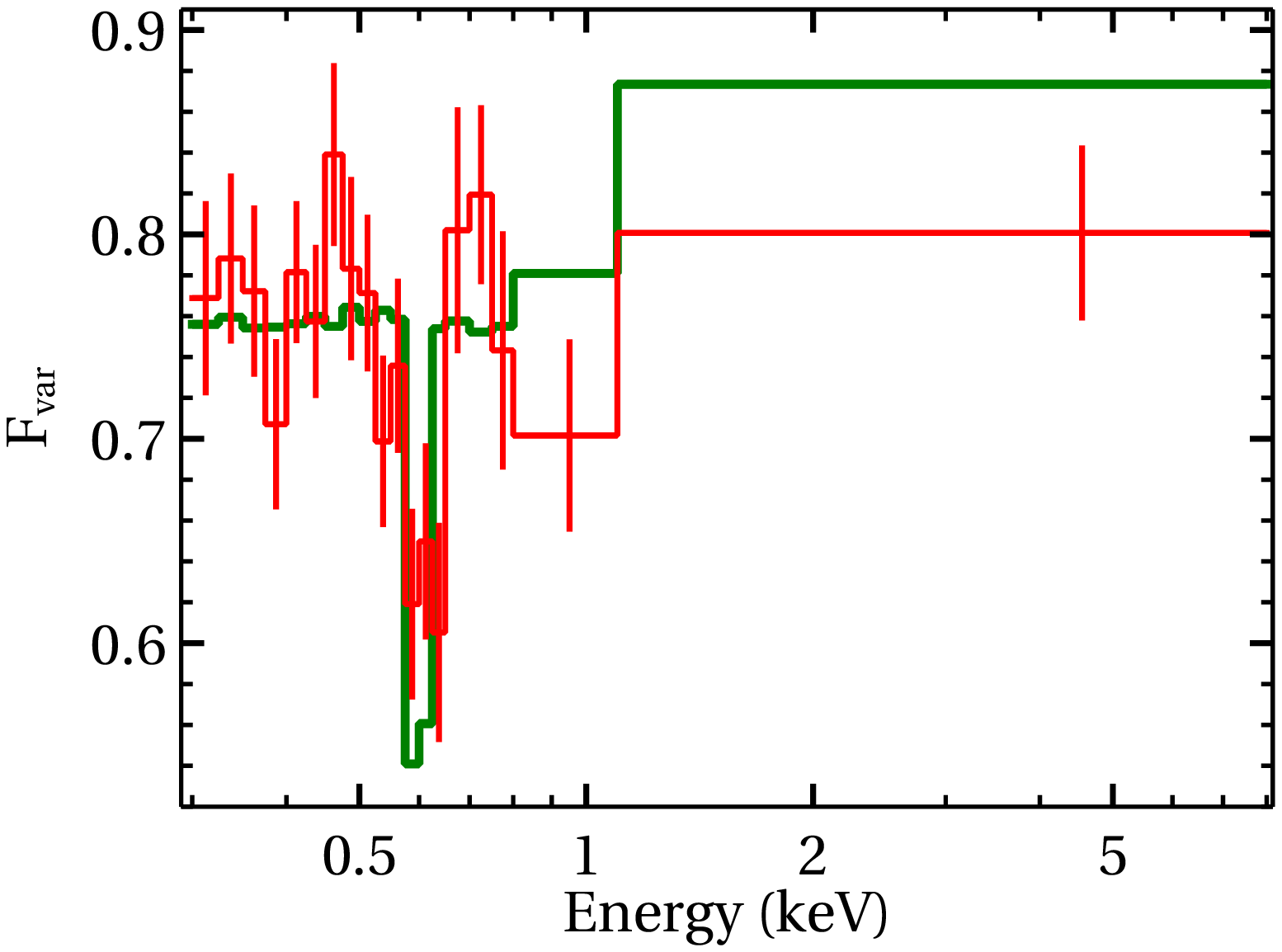}
\includegraphics[scale=0.5,angle=-0]{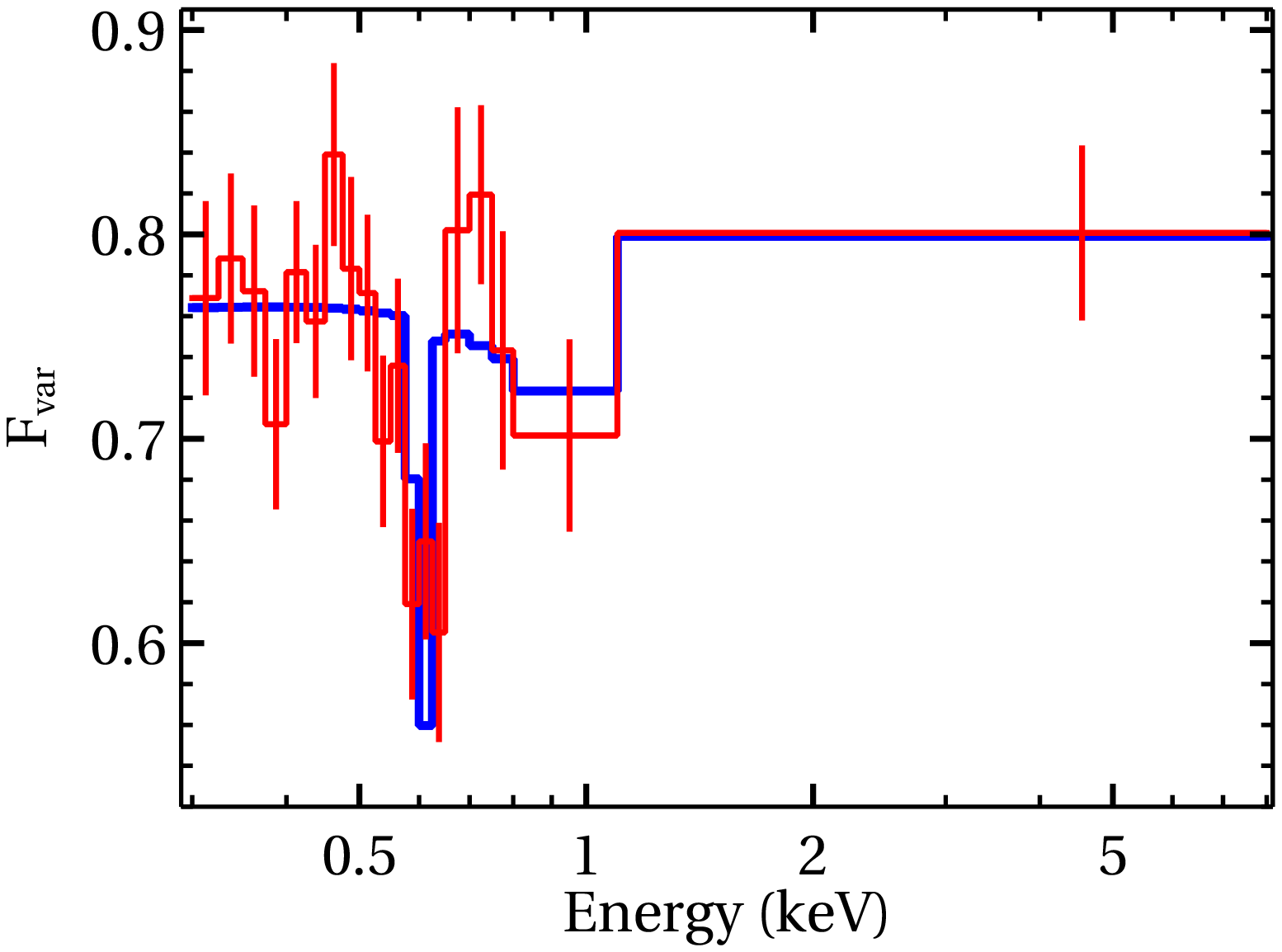}
\caption{Left: The 0.3$-$8\keV{} fractional rms spectrum and the `one-component disk Comptonization' model (solid green) in which the 0.6\keV{} emission line ({\tt{GL}}) is constant and the disk Comptonization ({\tt{optxagnf}}) is variable in luminosity only. Right: The 0.3$-$8\keV{} fractional rms spectrum and the best-fit `two-component relativistic reflection' model (solid blue) where both inner disk reflection ({\tt{relconv$\ast$reflionx}}) and illuminating continuum ({\tt{nthcomp}}) are variable in normalization and perfectly correlated with each other. The 0.6\keV{} emission line ({\tt{GL}}) is constant and hence shows a drop in variability at $\sim0.6$\keV{}.}
\label{fig7b}
\end{center}
\end{figure*}

\section{Fractional rms Spectral Modeling}
\label{sec:fvar}
To estimate the percentage of variability in the primary power-law continuum and soft X-ray excess emission, and also to quantify the variability relation between them, we derived and modeled the fractional rms variability spectrum of PG~1404$+$226. First, we extracted the background-subtracted, deadtime-corrected light curves in 19 different energy bands from the simultaneous and equal length ($72\ks{}$) combined EPIC-PN+MOS data with a time resolution of $\Delta t=500$\s{}. We have chosen the energy bands so that the minimum average count in each bin is around 20. Then we computed the frequency-averaged ($\nu\sim$[$1.4-100$]$\times10^{-5}$\hz{}) fractional rms, $F_{\rm{var}}$ in each light curve using the method described in \citet{va03}. We show the derived fractional rms spectrum of PG~1404$+$226 in Figure~\ref{fig7a} (left). The shape of the spectrum is approximately constant with a sharp drop at around 0.6\keV{}, which can be explained in the framework of a non-variable emission line component at $\sim0.6$\keV{} and two variable spectral components: the soft X-ray excess and primary power-law emission with the decreasing relative importance of the soft excess emission and increasing dominance of the primary power-law emission with energy. We constructed fractional rms spectral models using our best-fit phenomenological and physical mean spectral models in {\tt{ISIS}}~v.1.6.2-40 \citep{ho00}. 

First, we explored the phenomenological fractional rms spectral model in which the observed $0.6$\keV{} Gaussian emission line ({\tt{GL}}) is non-variable, and both the soft X-ray excess ({\tt{zbbody}}) and primary power-law emission ({\tt{zpowerlw}}) are variable in normalization and correlated with each other. Using the equation~(3) of \citet{ma17}, we obtained the expression for the fractional rms spectral model: 

\begin{footnotesize}
\begin{equation}
 F_{\rm{var}}=\frac{\sqrt{\big[(\frac{\Delta A_{\rm PL}}{A_{\rm PL}}f_{\rm PL})^{2}+(\frac{\Delta A_{\rm BB}}{A_{\rm BB}}f_{\rm BB})^{2}+2\gamma\frac{\Delta A_{\rm PL}}{A_{\rm PL}}\frac{\Delta A_{\rm BB}}{A_{\rm BB}}f_{\rm PL}f_{\rm BB}\big]}}{f_{\rm PL}(A_{\rm PL},E)+f_{\rm BB}(A_{\rm BB},E)+f_{\rm GL}(E)}  
\label{eu1}
\end{equation}
\end{footnotesize}
where $\frac{\Delta A_{\rm PL}}{A_{\rm PL}}$ and $\frac{\Delta A_{\rm BB}}{A_{\rm BB}}$ represent fractional changes in the normalization of the primary power-law, $f_{\rm PL}$ and blackbody, $f_{\rm BB}$ components respectively. $\gamma$ measures the correlation or coupling between $f_{\rm PL}$ and $f_{\rm BB}$. $f_{\rm GL}(E)$ represents the Gaussian emission line component with the observed energy of $\sim0.6$\keV{}.

We then fitted the 0.3$-$8\keV{} fractional rms spectrum of PG~1404$+$226 using this `two-component phenomenological' model (equation~\ref{eu1}) and the best-fit mean spectral model parameters as the input parameters for the above model. This model describes the data reasonably well with $\chi^{2}$/d.o.f = 25/16. The best-fit rms model parameters are: $\frac{\Delta A_{\rm PL}}{A_{\rm PL}}=0.8\pm0.7$, $\frac{\Delta A_{\rm BB}}{A_{\rm BB}}=0.78\pm0.03$ and $\gamma=0.68^{+0.32}_{-0.43}$. We show the fractional rms variability spectrum and the best-fit `two-component phenomenological' model in Fig.~\ref{fig7a} (right).

In PG~1404$+$226, the soft X-ray excess emission was modeled by two different physical models: intrinsic disk Comptonization and relativistic reflection from the ionized accretion disk. To break the degeneracy between these two possible physical scenarios, we made fractional rms spectral model considering our best-fit disk Comptonization and relativistic reflection models to the time-averaged spectrum. 

In the disk Comptonization scenario, the observed variability in PG~1404$+$226 was driven by variation in the source luminosity, as inferred from the joint fitting of 5 EPIC-PN spectra. Therefore, we can write the expression for the fractional rms (see \citealt{ma16,ma17}) as 

\begin{small}
\begin{equation}
 F_{\rm var}=\frac{\sqrt{<(\Delta f(L,E))^2>}}{f_{\rm optx}(L,E)+f_{\rm GL}(E)}
 \label{eu2}
\end{equation} 
\end{small} 
where $f_{\rm optx}(L,E)$ and $f_{\rm GL}(E)$ represent the best-fit disk Comptonization ({\tt{optxagnf}}) and $0.6$\keV{} Gaussian emission line ({\tt{GL}}) components, respectively. $L$ is the source luminosity which is the only variable free parameter in the model. The fitting of the 0.3$-$8\keV{} fractional rms spectrum using this model (equation~\ref{eu2}) resulted in an enhanced variability in the hard band above $\sim1$\keV{} with $\chi^{2}$/d.o.f =32/18. We show the fractional rms variability spectrum and the `one-component disk Comptonization' model in Figure~\ref{fig7b}~(left).

Then, we investigated the relativistic reflection scenario where the origin of the soft X-ray excess emission was explained with the disk irradiation \citep{cr05}. In this scenario, the rapid X-ray variability in PG~1404$+$226 can be described due to changes in the normalization of the illuminating power-law continuum and reflected inner disk emission as evident from the time-resolved spectroscopy. Thus, we constructed the `two-component relativistic reflection' model where both the inner disk reflection ({\tt{relconv$\ast$reflionx}}) and illuminating continuum ({\tt{nthcomp}}) are variable in normalization and perfectly correlated with each other. Mathematically, we can write the expression for the fractional rms as
\begin{small}
\begin{equation}
 F_{\rm var}=\frac{\sqrt{<(\Delta f(A_{\rm NTH},A_{\rm REF},E))^2>}}{f(A,E)}
  \label{eu3}
\end{equation}
\end{small}
where 
\begin{small}
\begin{equation}
f(A_{\rm NTH},A_{\rm REF},E)=f_{\rm NTH}(A_{\rm NTH},E)+f_{\rm REF}(A_{\rm REF},E)+f_{\rm GL}(E)
\end{equation}
\end{small}
Here $f_{\rm NTH}(A_{\rm NTH},E)$, $f_{\rm REF}(A_{\rm REF},E)$ and $f_{\rm GL}(E)$ represent the best-fit illuminating continuum ({\tt{nthcomp}}), inner disk reflection ({\tt{relconv$\ast$reflionx}}) and the $0.6$\keV{} Gaussian emission line ({\tt{GL}}) components, respectively. The two variable free parameters of this model (equation~\ref{eu3}) are $A_{\rm NTH}$ and $A_{\rm REF}$. We then fitted the observed fractional rms spectrum using the `two-component relativistic reflection' model which describes the data well with $\chi^{2}$/d.o.f =25/17. We show the fractional rms variability spectrum and the best-fit model in Figure~\ref{fig7b}~(right). The fractional variations in the normalization of the illuminating continuum and reflected emission are $\frac{\Delta A_{\rm NTH}}{A_{\rm NTH}}=0.83^{+0.17}_{-0.20}$ and $\frac{\Delta A_{\rm REF}}{A_{\rm REF}}=0.78^{+0.02}_{-0.03}$, respectively.

\section{Summary and Discussion}
\label{sec:discussion}
We present the first results from our \xmm{} observation of the NLS1 galaxy PG~1404$+$226. Here, investigate the large-amplitude X-ray variability, the origin of the soft X-ray excess emission and its connection with the intrinsic power-law emission through a detailed analysis of the time-averaged as well as time-resolved X-ray spectra, and frequency-averaged ($\nu\sim$[$1.4-100$]$\times10^{-5}$\hz{}) X-ray fractional rms spectrum. Below we summarize our results:

\begin{enumerate}
\item PG~1404$+$226 showed a short-term, large-amplitude variability event in which the X-ray (0.3$-$8\keV{}) count rate increased exponentially by a factor of $\sim7$ in about 10\ks{} and dropped sharply during the 2016 \xmm{} observation. The hard X-ray (1$-$8\keV{}) \chan{}/ACIS light curve also showed a rapid variability (a factor of $\sim2$ in about 5\ks{}) with an exponential rise and a sharp fall in 2000 \citep{da05}. The rapid X-ray variability had been observed in a few NLS1 galaxies (e.g. NGC~4051: \citealt{gi06}, 1H~0707--495: \citealt{fa12}, Mrk~335: \citealt{wi15}). However, the UV ($\lambda_{\rm eff}=2910\textrm{\AA}$) emission from PG~1404$+$226 is much less variable ($F_{\rm var,UV}\sim1\%$) compared to the X-ray (0.3$-$8\keV{}) variability ($F_{\rm var, X}\sim82\%$). 

\item The source exhibited strong soft X-ray excess emission below $\sim1$\keV{}, which was fitted by both the intrinsic disk Comptonization and relativistic reflection models. The EPIC-PN spectral data revealed the presence of a highly ionized ($\xi\sim600$~erg~cm~s$^{-1}$) Ne~X Lyman-$\alpha$ absorbing cloud along the line-of-sight with a column density of $N_{\rm H}\sim5\times10^{22}$~cm$^{-2}$ and a possible O~VIII Lyman-$\alpha$ emission line. However, we did not detect the presence of any outflow as found by \citet{da05}.

\item The modelling of the RGS spectrum not only confirms the presence of the Ne~X Lyman-$\alpha$ absorbing cloud and O~VIII Lyman-$\alpha$ emission line but also reveals an O~VIII Lyman-$\beta$ emission line.

\item The time-resolved spectroscopy showed a significant variability both in the soft X-ray excess and primary power-law flux, although there were no noticeable variations in the soft X-ray excess temperature ($kT_{\rm SE}\sim100$\ev{}) and photon index of the primary power-law continuum. 

\item In the disk Comptonization scenario, the rapid X-ray variability can be attributed to a variation in the source luminosity as indicated by the time-resolved spectroscopy. However, the modeling of the X-ray fractional rms spectrum using the `one-component disk Comptonization' model cannot reproduce the observed hard X-ray variability pattern and indicates reflection origin for the soft X-ray excess emission (see Fig.~\ref{fig7b}, left). 

\item In the relativistic reflection scenario, the observed large-amplitude X-ray variability was predominantly due to two components: illuminating continuum and smeared reflected emission, both of them are variable in normalization (see Fig.~\ref{fig7b}, right). 

\item The inner disk radius and central black hole spin as estimated from the relativistic reflection model are $r_{\rm in}<1.7 r_{\rm g}$ and $a>0.992$, respectively. \citet{cr05} also showed that the disk reflection could successfully explain the broadband (0.3$-$8\keV{}) spectrum of PG~1404$+$226 with the radiation from the inner accretion disk around a Kerr black hole. The disk inclination angle estimated from the ionized reflection model is $i^{\circ}=56.8^{+1.8}_{-12.9}$ which is in close agreement with that ($i^{\circ}=58^{+7}_{-34}$) obtained by \citet{cr05}. The non-detection of the 6.4\keV{} iron emission line could be due to its smearing on the broad shape in the spectrum.

\item We found that the soft (0.3$-$1\keV{}) and hard (1$-$8\keV{}) band count rates are correlated with each other and vary in a non-linear manner as suggested by the steepening of the flux-flux plot. The fitting of the hard-vs-soft counts plot with a power-law plus constant (PLC) model reveals a significant positive offset at high energies which can be interpreted as corroboration for the presence of a less variable reflection component (probably smeared iron emission line) in the hard band on timescales of $\sim20$~hr.  

\end{enumerate}

\subsection{UV/X-ray Variability and Origin of the Soft X-ray Excess Emission}

The observed UV variability in PG~1404$+$226 is weak with $F_{\rm var}\sim1$~per~cent only, whereas the X-ray variability is much stronger ($F_{\rm var}\sim82$~per~cent) on timescales of $\sim73$\ks{}. The UV and soft X-ray excess emission do not occur to be significantly correlated as demonstrated in Fig.~\ref{fig6}~(Right). 

In the intrinsic disk Comptonization ({\tt{optxagnf}}) model, the soft X-ray excess emission results from the Compton up-scattering of the UV seed photons by an optically thick, warm ($kT_{\rm SE}\sim0.1-0.2$\keV{}) electron plasma in the inner disk (below $r_{\rm corona}$) itself. So, if the soft X-ray excess was the direct thermal emission from the inner accretion disk, then we expect correlated UV/soft excess variability. However, we did not find any correlation between the UV flux and X-ray spectral parameters. Furthermore, the modeling of the rms variability spectrum using `one-component disk Comptonization' model could not describe the observed hard X-ray variability in PG~1404$+$226 (see Fig.~\ref{fig7b}, left). It might be possible that the UV and X-ray emitting regions interact on a timescale much longer than the duration of our observation. To explore that possibility, we calculated various timescales associated with the accretion disk. The light travel time between the central X-ray source and the standard accretion disk is given by the relation \citep{de15}

\begin{small}
\begin{equation}
t_{\rm cross}=2.6\times10^{5}\left(\frac{\lambda_{\rm eff}}{3000\textrm{\AA}}\right)^{4/3}\left(\frac{\dot{M}}{\dot{M_{\rm E}}}\right)^{1/3}\left(\frac{M_{\rm BH}}{10^{8}M_{\odot}}\right)^{2/3}~\s{}
\label{eu4}
\end{equation}
\end{small}
where $\frac{\dot{M}}{\dot{M_{\rm E}}}$ is the scaled mass accretion rate, $M_{\rm BH}$ is the central black hole mass in units of $M_{\odot}$ and $\lambda_{\rm eff}$ is the effective wavelength where the disk emission peaks. 
In the case of PG~1404$+$226, $M_{\rm BH}\sim4.5\times10^6M_{\odot}$, $\frac{\dot{M}}{\dot{M_{\rm E}}}\sim0.08$ (as obtained from the {\tt{optxagnf}} model as well as calculated from the unabsorbed flux in the energy band 0.001$-$100\keV{} using the convolution model {\tt cflux} in {\tt XSPEC}) and $\lambda_{\rm eff}=2910\textrm{\AA}$ for UVW1 filter. Therefore, the light crossing time between the X-ray source and the disk is $\sim13.6$\ks{}, which corresponds to the peak disk emission radius of $\sim600r_{\rm g}$. If we consider a thin disk for which height-to-radius ratio, $h/r\sim0.1$ \citep{cz06}, the viscous timescale at this emission radius ($r\sim600r_{\rm g}$) is of the order of $\sim10$~years which is much longer than the time span of our \xmm{} observation. Although both the soft and hard X-ray emission from PG~1404$+$226 are highly variable, the lack of any strong UV variability is in contradiction with the viscous propagation fluctuation scenario. 

In the relativistic reflection model, the soft X-ray excess is a consequence of disk irradiation by a hot, compact corona close to the black hole. We found a strong correlation between the soft and hard X-ray emission which is expected in the reflection scenario. Additionally, the modeling of the fractional rms spectrum considering `two-component relativistic reflection' model can reproduce the observed X-ray variability very well (see Fig.~\ref{fig7b}, Right). 

\subsection{Origin of Rapid X-ray Variability}
PG~1404$+$226 shows a strong X-ray variability with the fractional rms amplitude of $F_{\rm var, X}\sim82\%$ on timescales of $\sim20$~hr. We attempted to explain the observed rapid variability of PG~1404$+$226 in the framework of two possible physical scenarios: intrinsic disk Comptonization and relativistic reflection from the ionized accretion disk. In the disk Comptonization scenario, if the rapid X-ray variability is due to the variation in the source luminosity which is favored by the time-resolved spectroscopy, then it slightly overpredicts the fraction variability in the hard band (see Fig.~\ref{fig7b}, left). Therefore, it is unlikely that the rapid X-ray variability is caused by variations in the source (warm plus hot coronae) luminosity only. On the other hand, the soft X-ray excess and primary continuum vary non-linearly (as $F_{{\rm primary}}\propto F_{{\rm excess}}^{1.54}$), which indicates that the soft X-ray excess is reciprocating with the primary continuum variations, albeit with a smaller amplitude. It is in agreement with the smeared reflection scenario which is further supported by the high emissivity index ($q\sim9.9$) and non-detection of the iron line. Moreover, the fractional variability spectrum of PG~1404$+$226 is best described by two components: illuminating continuum and reflected emission, both of them are variable in flux (see Fig.~\ref{fig7b}, right). We interpret these rapid variations in the framework of the light bending model \citep{mi03,mi04,mf04,fa05}, according to which the primary coronal emission is bent down onto the accretion disk due to strong gravity and forms reflection components including the soft X-ray excess emission. The nature of the rapid X-ray variability in PG~1404$+$226 prefers the `lamppost geometry' for the primary X-ray emitting hot corona.

\acknowledgments
LM gratefully acknowledges support from the University Grants Commission (UGC), Government of India. The authors thank the anonymous referee for constructive suggestions in improving the quality of the paper. This research has made use of processed data of \xmm{} observatory through the High Energy Astrophysics Science Archive Research Center Online Service, provided by the NASA Goddard Space Flight Center. This research has made use of the NASA/IPAC Extragalactic Database (NED), which is operated by the Jet Propulsion Laboratory, California Institute of Technology, under contract with the NASA. Figures in this manuscript were made with the graphics package \textsc{pgplot} and GUI scientific plotting package \textsc{veusz}. 

\facility{\xmm{} (EPIC-PN, EPIC-MOS, RGS and OM)}

\software{HEASOFT, SAS, XSPEC \citep{ar96}, XSTAR \citep{kb01}, FTOOLS \citep{bl95}, ISIS \citep{ho00}, Python, S-Lang, Veusz, PGPLOT}

\label{lastpage}
\end{document}